\documentclass[twocolumn,tighten]{aastex63}

\shorttitle{Spectral Index Imaging for the ngEHT}
\shortauthors{Chael et al.}
\graphicspath{{./}{figures/}}

\usepackage{amsmath,mathtools,mathrsfs,amssymb}
\usepackage{relsize}
\usepackage{hyperref}
\usepackage{verbatim}
\usepackage{xspace}
\usepackage{graphicx}
\usepackage{longtable}
\usepackage{newtxtext,newtxmath}
\usepackage[T1]{fontenc}
\usepackage{xspace}

\def\d{\partial}
\def\uv{$u-v$\xspace}

\def\dvec{\mathbf{d}}

\def\i{\mathrm{i}}
\def\dvec{\mathbf{d}}

\def\lsim{\mathrel{\raise.3ex\hbox{$<$\kern-.75em\lower1ex\hbox{$\sim$}}}}
\def\gsim{\mathrel{\raise.3ex\hbox{$>$\kern-.75em\lower1ex\hbox{$\sim$}}}}
\def\gtwid{\mathrel{\raise.3ex\hbox{$>$\kern-.75em\lower1ex\hbox{$\sim$}}}}
\def\proptwid{\mathrel{\raise.3ex\hbox{$\propto$\kern-.75em\lower1ex\hbox{$\sim$}}}}

\DeclareRobustCommand{\okina}{%
  \raisebox{\dimexpr\fontcharht\font`A-\height}{%
    \scalebox{0.8}{`}%
  }%
}

\begin{document}

\title{Multi-frequency Black Hole Imaging for the Next-Generation Event Horizon Telescope}

\correspondingauthor{Andrew Chael}
\email{achael@princeton.edu}

\author[0000-0003-2966-6220]{Andrew Chael}
\altaffiliation{NASA Hubble Fellowship Program, Einstein Fellow}
\affiliation{Princeton Gravity Initiative, Princeton University, Jadwin Hall, Princeton, NJ 08544, USA}

\author[0000-0002-5297-921X]{Sara Issaoun}
\altaffiliation{NASA Hubble Fellowship Program, Einstein Fellow}
\affiliation{Center for Astrophysics $|$ Harvard \& Smithsonian, 60 Garden Street, Cambridge, MA 02138, USA}

\author[0000-0002-5278-9221]{Dominic W. Pesce}
\affiliation{Center for Astrophysics $|$ Harvard \& Smithsonian, 60 Garden Street, Cambridge, MA 02138, USA}
\affiliation{Black Hole Initiative at Harvard University, 20 Garden Street, Cambridge, MA 02138, USA}

\author[0000-0002-4120-3029]{Michael D. Johnson}
\affiliation{Center for Astrophysics $|$ Harvard \& Smithsonian, 60 Garden Street, Cambridge, MA 02138, USA}
\affiliation{Black Hole Initiative at Harvard University, 20 Garden Street, Cambridge, MA 02138, USA}

\author[0000-0001-5287-0452]{Angelo Ricarte}
\affiliation{Black Hole Initiative at Harvard University, 20 Garden Street, Cambridge, MA 02138, USA}
\affiliation{Center for Astrophysics $|$ Harvard \& Smithsonian, 60 Garden Street, Cambridge, MA 02138, USA}

\author[0000-0002-1827-1656]{Christian M. Fromm}
\affiliation{Institut f\"ur Theoretische Physik und Astrophysik, Universit\"at W\"urzburg, Emil-Fischer-Strasse 31, 97074
W\"urzburg, Germany}
\affiliation{Institut f\"ur Theoretische Physik, Goethe Universit\"at, Max-von-Laue-Str. 1, D-60438 Frankfurt, Germany}
\affiliation{Max-Planck-Institut f\"ur Radioastronomie, Auf dem H\"ugel 69, D-53121 Bonn, Germany}

\author[0000-0002-8131-6730]{Yosuke Mizuno}
\affiliation{Tsung-Dao Lee Institute, Shanghai Jiao Tong University, 520 Shengrong Road, Shanghai, 201210, China}
\affiliation{School of Physics and Astronomy, Shanghai Jiao Tong University, 800 Dongchuan Road, Shanghai, 200240, China}
\affiliation{Institut f\"ur Theoretische Physik, Goethe Universit\"at, Max-von-Laue-Str. 1, D-60438 Frankfurt, Germany}


\begin{abstract}
The Event Horizon Telescope (EHT) has produced images of the plasma flow around the supermassive black holes in Sgr A* and M87* with a resolution comparable to the projected size of their event horizons. Observations with the next-generation Event Horizon Telescope (ngEHT) will have significantly improved Fourier plane coverage and will be conducted at multiple frequency bands (86, 230, and 345 GHz), each with a wide bandwidth. At these frequencies, both Sgr A* and M87* transition from optically thin to optically thick. Resolved spectral index maps in the near-horizon and jet-launching regions of these supermassive black hole sources can constrain properties of the emitting plasma that are degenerate in single-frequency images. In addition, combining information from data obtained at multiple frequencies is a powerful tool for interferometric image reconstruction, since gaps in spatial scales in single-frequency observations can be filled in with information from other frequencies. Here we present a new method of simultaneously reconstructing interferometric images at multiple frequencies along with their spectral index maps. The method is based on existing Regularized Maximum Likelihood (RML) methods commonly used for EHT imaging and is implemented in the  \texttt{eht-imaging} Python software library. We show results of this method on simulated ngEHT data sets as well as on real data from the VLBA and ALMA. These examples demonstrate that simultaneous RML multi-frequency image reconstruction produces higher-quality and more scientifically useful results than is possible from combining independent image reconstructions at each frequency.
\end{abstract}

\nocite{PaperI}
\nocite{PaperII}
\nocite{PaperIII}
\nocite{PaperIV}
\nocite{PaperV}
\nocite{PaperVI}
\nocite{PaperVII}
\nocite{PaperVIII}
\nocite{SgrAPaperI}
\nocite{SgrAPaperII}
\nocite{SgrAPaperIII}
\nocite{SgrAPaperIV}
\nocite{SgrAPaperV}
\nocite{SgrAPaperVI}
\nocite{SgrAPaperVII}
\nocite{SgrAPaperVIII}

\section{Introduction}
\label{sec:intro}

The Event Horizon Telescope (EHT) is a Very-Long-Baseline-Interferometry (VLBI) array operating at 1.3 mm wavelength (230 GHz) with a nominal resolution of $\approx\!20\,\mu$as  \citep{PaperII}. Recent EHT observations \citep{PaperIII,SgrAPaperII} have produced the first images of two supermassive black hole sources with resolution on the scale of their event horizons; in M87*
\citep[$M\approx6.5\times 10^9 M_\odot$;][]{PaperI} and Sgr~A* \citep[$M\approx4\times10^6 M_\odot$;][]{SgrAPaperI}.
In both Sgr A* and M87*, the 230 GHz EHT images feature a ring with a diameter approximately five times the projected Schwarzschild radius, consistent with the predicted size of the ``black hole shadow'' \citep{PaperIV,SgrAPaperIII}. In both sources, EHT total intensity images constrain the black hole mass \citep{PaperVI,SgrAPaperIV}, provide tests of the Kerr metric \citep{PaperVI,SgrAPaperVI}, and constrain potential scenarios for the physics of accretion and jet launching around black holes \citep{PaperV,SgrAPaperV}. Linearly polarized EHT images of M87* \citep{PaperVII} more strongly constrain the nature of the accretion flow and jet; they indicate that magnetic fields around M87* are strong and dynamically important \citep{PaperVIII}.

Building on the success of these observations, the proposed next-generation Event Horizon Telescope (ngEHT) plans to add $\sim$10 new telescopes to the EHT. These additional sites will fill in the EHT's sparse \uv plane coverage, enhancing imaging dynamic range and enabling the recovery of faint features in the M87* jet. Rapid filling of the \uv plane from these additional sites will also allow for robust imaging of rapid variability in the accretion flow around Sgr A*. In addition to adding new sites, the ngEHT will also increase the observing bandwidth and the range of observed frequencies up to 345 GHz and (potentially) down to 86 GHz \citep{Doeleman_2019, Issaoun_2022}.

EHT observations of Sgr A* and M87* have not yet produced a spatially resolved map of spectral index, the slope of the image log-intensity as a function of log-frequency.  The relatively narrow 4 GHz total bandwidth of 2017 EHT observations at 1.3 mm and a lack of simultaneous observations at lower frequencies has prevented the creation of spectral index maps from EHT data thus far.
Future ngEHT datasets with simultaneous, wide-bandwidth observations at  86, 230, and 345~GHz
may enable the robust recovery of resolved spectral index maps, which will be crucial for constraining the physics of the accretion flow and jet-launching regions near these supermassive black holes. \citet{Ricarte_2022} studied spectral index maps of horizon-scale models of both M87* and Sgr A* and explored how the spectral index encodes information on the plasma density, temperature, magnetic field strength, and electron distribution function.  These plasma parameters can vary by orders of magnitude among General Relativistic Magnetohydrodynamic (GRMHD) simulation models that all produce similar 230~GHz images of these sources; spectral index maps can thus help break degeneracies between physical parameters and test models in unprecedented detail.

Simultaneous imaging of interferometric data separated in frequency is not new to interferometry. The CLEAN algorithm \citep{Hog_1974} is the standard choice for reconstructing interferometric images from data. Several CLEAN-based algorithms that simultaneously use observations at multiple frequencies to solve for an image and spectral index map have been developed \citep{Sault_1994,Rau_2011,Offringa_2021} and are implemented in standard interferometric software.\footnote{e.g. CASA \texttt{tclean}, \url{https://casadocs.readthedocs.io/en/stable/api/tt/casatasks.imaging.tclean.html}}
The imaging method we present in this paper is fundamentally different from these CLEAN-based approaches, however.
As a Regularized Maximum Likelihood (RML) image reconstruction method \citep[e.g.][]{Akiyama_2016,Chael18,PaperIV}, rather than deconvolving point-spread function structure from the multi-frequency images after a Fourier transform of the gridded data,
we forward-model the observed data from an underlying image model and attempt to find the best fit to the data subject to prior constraints. Our method for multi-frequency RML imaging is a straightforward adaptation of single-frequency RML imaging commonly used for EHT image reconstruction; critically, it allows us to maintain the flexibility of the RML approach in fitting directly to interferometric `closure quantities', which are robust to systematic calibration uncertainties that affect all millimeter-VLBI datasets.

In this paper, we present a new method for multi-frequency imaging of interferometric data. The method is a simple extension of the RML approach commonly applied to EHT data sets. It combines heterogeneous data products observed at multiple frequencies in a simultaneous fit of a reference frequency image, spectral index map, and potentially a spectral curvature map.  In \autoref{sec:method}, we review the RML method for interferometric imaging and present its extension to multi-frequency imaging used in this paper. In \autoref{sec:Implementation}, we discuss the implementation of spectral index imaging in the \texttt{eht-imaging} software package. In \autoref{sec:ngEHT}, we present results of applying the method to simulated ngEHT datasets of different models of M87*. In \autoref{sec:realdata}, we present results of applying the method to real observations from the VLBA and ALMA. In \autoref{sec:Discussion} we discuss the results and potential extensions to the method presented here, and we conclude in \autoref{sec:Conclusion}.

\section{Method}
\label{sec:method}

\subsection{Imaging with Regularized Maximum Likelihood (RML)}
\label{sec:rml}

VLBI arrays like the EHT and ngEHT measure complex visibilities ($V_{\rm ab}$) between two widely-separated radio telescopes $a$ and $b$ by correlating the time-series of electric field values recorded at each location. In the ideal case, this visibility $V_{\rm ab}$ is a measurement of the Fourier transform of the source image intensity distribution $I(x,y)$:
\begin{align}
 \label{eq::VCZ}
 V_{\rm ab}= \tilde{I}(u,v) = \int\int\, I(x,y)\,e^{-2\pi \mathrm{i} (ux+vy)}\, \mathrm{d}x\,\mathrm{d}y,
\end{align}
where $x$ and $y$ are angular coordinates on the sky, and $u$ and $v$ are the coordinates of the projected baseline $\vec{b}_{\rm ab}$ between the sites, measured in wavelengths.

Most methods for reconstructing images from VLBI measurements use the CLEAN algorithm \citep{Hog_1974, Clark_80}. CLEAN makes use of the Fourier relationship between measured complex visibilities $V_{\rm ab}$ and the source intensity distribution $I(x,y)$  by first constructing the ``dirty image''  from the inverse Fourier transform of the sparse set of measured complex visibilities, with all unmeasured visibilities set to zero. The CLEAN algorithm then works in the image plane by iteratively deconvolving the interferometer ``dirty beam,'' or point-spread response function, from the dirty image to obtain a model image for the underlying source.

By contrast, an increasingly utilized approach for reconstructing VLBI images from data is the class of RML algorithms. Instead of attempting to de-convolve artifacts from an image in the source plane, RML methods forward-model an interferometric dataset $\mathbf{d}$ from a trial image $\mathbf{I}$ and iteratively adjust this image to find the best-fit to the data while also satisfying  prior constraints on the image structure. RML methods find the best-fit image by minimizing an objective function $J(\mathbf{I})$ composed of log-likelihood terms $\mathcal{L}\left(\mathbf{d} | \mathbf{I}\right)$ and ``regularizer'' terms $S_R\left(\mathbf{I}\right)$, which favor or penalize certain image features \citep[see, e.g.,][]{Chael18_Closure,PaperIV}:
\begin{equation}
 \label{eq::objfunc}
 J\left(\mathbf{I}\right) = -\mathlarger{\sum}_{\mathclap{\text{data terms}}} \kappa_D \log \mathcal{L}\left(\mathbf{d} | \mathbf{I}\right) \;\; - \mathlarger{\sum}_{\mathclap{\text{regularizers}}} \lambda_R S_R\left(\mathbf{I}\right).
\end{equation}
In \autoref{eq::objfunc}, the $\kappa_R$ and $\lambda_D$ terms are hyperparameters that set the relative weight of the different likelihood and regularizer terms in the total functional $J$ being minimized. For specific combinations of likelihood terms ($\mathcal{L}$), regularizer terms ($S_R$), and hyperparameters ($\kappa_D,\lambda_R$), the RML objective function \autoref{eq::objfunc} can be equivalent to a log-posterior probability distribution, where $\lambda_R S_R(\mathbf{I})$ is the log-prior. In general, however, RML methods freely use combinations of log-likelihoods, regularizer terms, and hyperparameters; the objective function $J$ \autoref{eq::objfunc} usually is not a rigorously-defined log-posterior.

Commonly used regularizer terms $S(\mathbf{I})$ for imaging EHT data sets include image entropy (i.e. the Maximum Entropy Method, \citealt{Frieden_1972, GS_1978, Cornwell_1985, NN_1986}), an $\ell_1$ norm term to promote spatial sparsity \citep[e.g.][]{Honma_2014,Akiyama_bs}, and image smoothness terms like total variation \citep[e.g.][]{TV, Thiebaut_2017} or total squared variation \citep[e.g.][]{kuramochi_tv2}. A full list of the image regularizers implemented in the \texttt{eht-imaging} RML code and used in this paper can be found in \citet{Chael18_Closure} and \citet{PaperIV}.

\subsection{RML imaging across frequency}
\label{sec:rmlmf}

In this paper we perform a simple extension of the RML method described above to simultaneously reconstruct images from data obtained at a set of frequencies $\{\nu_i\}$.
We denote observational data products obtained at each frequency $\nu_i$ by $\dvec_i$. These data products at each frequency can be heterogeneous: they may consist of calibrated complex visibilities $V$, visibility amplitudes $|V|$, closure phases $\psi_\mathcal{C}$, closure amplitudes $A_\mathcal{C}$, or any other product derivable from the interferometric data. Our goal is to take this set of data $\{\mathbf{d}_i\}$ and reconstruct the images at each frequency $\mathbf{I}_{i}$ that best fit both the data and our regularizing assumptions on the structure of the image and the evolution of the image with frequency.

Instead of reconstructing images $\mathbf{I}_{i}$ at each frequency independently, we instead parameterize the images at each frequency with a log-log Taylor expansion around a reference image $\mathbf{I}_0$ at a chosen reference frequency $\nu_0$:
\begin{align}
    \log \mathbf{I}_{i} = \log\left[\mathbf{I}_0\right] + \boldsymbol{\alpha}\log\left(\frac{\nu_i}{\nu_0}\right)  + \boldsymbol{\beta}\log^2\left(\frac{\nu_i}{\nu_0}\right)  + ... 
\label{eq:loglogex}
\end{align}
In \autoref{eq:loglogex}, $\boldsymbol\alpha$ is the spectral index map and $\boldsymbol\beta$ is the spectral curvature map. In this paper we only work with these first two terms in the expansion, but in principle the same approach could be used to fit spectral terms at arbitrary order in the Taylor series around $\nu_0$.
If we define
\begin{align}
    x_i &= \log\left[\nu_i/\nu_0\right], \\
    \mathbf{y}_i &= \log\left[\mathbf{I}_i\right],
    \label{eq:transforms}
\end{align}
then we can summarize the expansion \autoref{eq:loglogex} to second order as
\begin{equation}
    \mathbf{I}_{i} = \text{exp}\left[\mathbf{y}_0 + \boldsymbol\alpha x_i + \boldsymbol\beta x^2_i\right].
    \label{eq:loglog2}
\end{equation}
We employ this image model (\autoref{eq:loglog2}) for all multi-frequency reconstructions in this paper. We occasionally solve only for the spectral index  $\boldsymbol\alpha$ and fix the curvature map $\boldsymbol\beta=0$.

After defining the image model, we simply extend the standard RML framework presented in \autoref{sec:rml} to produce simultaneous reconstructions of the reference frequency image $\mathbf{I}_{0}$, the spectral index map $\boldsymbol\alpha$, and potentially the curvature map $\boldsymbol\beta$. To do this, we minimize an expanded objective function:
\begin{align}
    J\left(\mathbf{I}_0,\boldsymbol\alpha,\boldsymbol\beta\right) = &\mathlarger\sum_{\text{data terms}}  \mathlarger\sum_{\nu_i} - \kappa_D \log \mathcal{L}\left(\mathbf{d}_i | \mathbf{I}_{i}\right) \;\; + \nonumber \\
    &\mathlarger\sum_{\rm regularizers} -\lambda_{I} S_{I}\left(\mathbf{I}_0\right)  - \lambda_{\alpha}S_{\alpha}\left(\boldsymbol\alpha\right) - \lambda_{\beta}S_{\beta}\left(\boldsymbol\beta\right).
\label{eq:objfuncI}
\end{align}
In \autoref{eq:objfuncI}, the $\mathcal{L}$ terms are log-likelihoods that compare the various data products $\mathbf{d}_i$ obtained at each frequency $\nu_i$ with the trial image reconstructions $\mathbf{I}_i$ at the same frequency. Each image $\mathbf{I}_i$ is a function of the underlying parameters $\left(\mathbf{I}_0,\boldsymbol\alpha,\boldsymbol\beta\right)$ through \autoref{eq:loglog2}. Critically, the RML method is flexible regarding data products; each frequency $\nu_i$ can have multiple log-likelihoods from different, heterogeneous data products that contribute to the total data constraint for that frequency.

The $S_I$ term in \autoref{eq:objfuncI} represents a regularizing term on the reference frequency image. This term can have multiple components, (e.g. maximum entropy, total variation,  $\ell_1$ norm).
Similarly, here we introduce $S_\alpha$ and $S_\beta$ regularizers in \autoref{eq:objfuncI} as regularizing terms on the spectral index map $\boldsymbol\alpha$ and curvature map $\boldsymbol\beta$.

We define regularizers on the resolved spectral index and curvature maps in close analogy with those already developed for total intensity RML imaging (\autoref{sec:rml}).
In this paper, we only consider two spectral regularizer terms. The first is an $\ell_2$ norm term that pushes the index $\alpha$ to a fiducial value in the absence of data constraints:
\begin{equation}
\label{eq:sell2}
    S_{\ell_2}(\boldsymbol\alpha) = -\frac{1}{N}\sum_{l,m} \left(\alpha_{lm} - \alpha_0\right)^2.
\end{equation}
In \autoref{eq:sell2}, the leading factor of the total number of pixels $N$ is a normalization factor that allows us to use similar hyperparameter values $\lambda_\alpha$ for different-sized images. The fiducial value $\alpha_0$ may be set to zero, or set to a measurement of the spectral index of the unresolved source. The same $\ell_2$ norm term in \autoref{eq:sell2} can be directly applied to regularize the spectral curvature map $\boldsymbol\beta$.

We also use a total variation regularizer on the spectral index/curvature maps to disfavor large variations in $\boldsymbol\alpha$ or $\boldsymbol\beta$ over small parts of the image:
\begin{align}
 S_{\text{TV}}(\boldsymbol\alpha) = -\frac{\Theta_{\rm beam}}{N\Delta\theta}\sum_{l,m} \left[ \left(\alpha_{l+1,m}-\alpha_{l,m}\right)^2 + \left(\alpha_{l,m+1}-\alpha_{l,m}\right)^2\right]^{1/2}.
 \label{eq:alphatv}
\end{align}
Again, in \autoref{eq:alphatv} the factor $\Theta_{\rm beam}/N\Delta\theta$ is a normalizing factor, where $\Theta_{\rm beam}$ is the size of the interferometer beam and $\Delta\theta$ is the pixel size \citep[cf.][Equation 35]{PaperIV}.  In this work, we only employ $S_{\rm TV}$ and $S_{\ell_2}$ regularizers for both the spectral index $\boldsymbol\alpha$ and spectral curvature $\boldsymbol\beta$ maps, but in principle, the multi-frequency generalization of RML imaging we present here can support many further regularizers on $\boldsymbol\alpha$ and $\boldsymbol\beta$.

In summary, the multi-frequency RML imaging procedure involves simultaneously solving for the image $\mathbf{I}_0$ at a reference frequency $\nu_0$ in addition to the spectral index and spectral curvature maps $\boldsymbol\alpha, \boldsymbol\beta$ by minimizing the objective function $J$ (\autoref{eq:objfuncI}). This procedure is a straightforward extension of the usual total intensity RML framework described in \autoref{sec:rml}.
It allows us to fit observations taken at many different reference frequencies or frequency channels simultaneously in an image model with a smaller number of degrees of freedom than in reconstructing images independently at each frequency; it also allows us to directly regularize the spectral index map.
As a result, it is also straightforward to implement RML multifrequency imaging in existing RML imaging codes and scripts developed for the EHT and other interferometric arrays \citep[e.g.][]{PaperIV}.

\begin{table*}[t]
    \centering
    \begin{tabular}{c|c|ccc}
    \hline
      Station Code  & Location  & 86 GHz & 230 GHz & 345 GHz  \\
      \hline \hline
      \textbf{Existing EHT sites} \\
      \hline \hline
       ALMA  & Antofagasta, Chile & {\bf X} & {\bf X} & {\bf X} \\
       APEX  & Antofagasta, Chile & - & {\bf X} & {\bf X} \\
       SMA  &Hawai\okina i, USA& - & {\bf X} & {\bf X} \\
       JCMT  &Hawai\okina i, USA& - & {\bf X} & {\bf X} \\
       LMT  &Puebla, Mexico& {\bf X} & {\bf X} & {\bf X} \\
       SMT  &Arizona, USA& - & {\bf X} & {\bf X} \\
       KP  &Arizona, USA& - & {\bf X} & - \\
       NOEMA  &Pr.-Alpes-C\^{o}te d'Azur, France& {\bf X} & {\bf X} & {\bf X} \\
       IRAM-30m (PV)  &Andalusia, Spain& {\bf X} & {\bf X} & {\bf X} \\
       SPT  &South Pole, Antarctica& - & {\bf X} & {\bf X} \\
       GLT  &Avannaata, Greenland& {\bf X} & {\bf X} & {\bf X} \\
      \hline \hline
      \textbf{Potential ngEHT sites} \\
      \hline \hline
       OVRO  &California, USA& {\bf X} & {\bf X} & {\bf X} \\
       BAJA  &Baja California, Mexico& {\bf X} & {\bf X} & {\bf X} \\
       BAR  &California, USA& {\bf X} & {\bf X} & {\bf X} \\
       CNI  &La Palma, Canary Islands, Spain& {\bf X} & {\bf X} & {\bf X} \\
       GAM  &Khomas, Namibia& {\bf X} & {\bf X} & -\\
       GARS  &Antarctic Peninsula, Antarctica& {\bf X} & {\bf X} & {\bf X} \\
       HAY  &Masachussetts, USA& {\bf X} & {\bf X} & {\bf X} \\
       NZ  &Canterbury, New Zealand& {\bf X} & {\bf X} & {\bf X} \\
       SGO  &Santiago, Chile & {\bf X} & {\bf X} & {\bf X} \\
       CAT  &R\'{i}o Negro, Argentina & {\bf X} & {\bf X} & {\bf X} \\
       \hline
    \end{tabular}
    \caption{Locations and frequency capabilities of active EHT stations and potential additional stations planned for the ngEHT. The geographical and weather properties of these sites were described in \citet{Raymond_2021}. The potential ngEHT station configuration used in this paper and its 86\,GHz capabilities are described in \citet{Issaoun_2022}.}
    \label{tab:ngEHTarray}
\end{table*}

\begin{figure*}[t!]
\centering
\includegraphics[width=\textwidth]{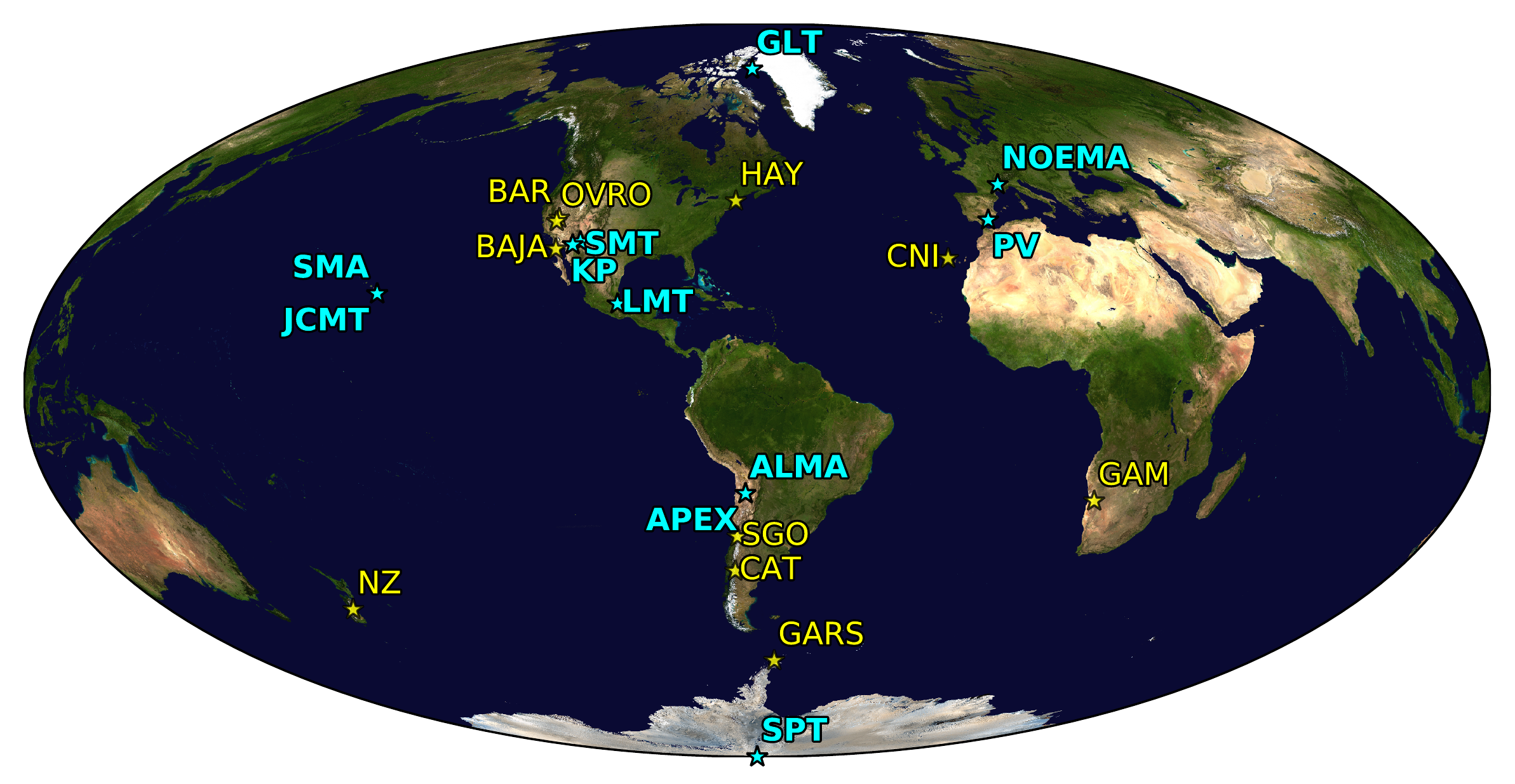}

\caption{A potential configuration of the ngEHT array, as described in \citet{Issaoun_2022}. The current EHT stations are shown in teal, while potential ngEHT stations are shown in yellow.}
\label{fig:ngEHT}
\end{figure*}

\section{Implementation in \texttt{eht-imaging}}
\label{sec:Implementation}

We implement the RML approach to multi-frequency synthesis described in \autoref{sec:method} in the \texttt{eht-imaging} Python software package  \citep{Chael_16,Chael18_Closure,ehtim}.\footnote{\url{https://github.com/achael/eht-imaging}}
The \texttt{eht-imaging} package provides routines for RML imaging of interferometric data in total intensity and polarization, synthetic data generation, visualization, and analytic model fitting and parameter exploration. The RML image reconstruction algorithms in \texttt{eht-imaging} have been used in the analysis of EHT observations of M87* \citep{PaperIV,PaperVIII}, Sgr A* \citep{SgrAPaperIII}, 3C279 \citep{Kim_2020} and Centaurus A \citep{Janssen_2021}. They have also been used in imaging VLBI data sets from other arrays at lower frequencies \citep[e.g.][]{Issaoun_2019,Issaoun_2021,Xu_2021,Savolainen_2021,Zhao_2022}.

\subsection{RML imaging data products}
A major advantage of RML methods for high-frequency VLBI arrays like the EHT and ngEHT is that while CLEAN-based methods require fully calibrated complex visibility measurements $V_{\rm ab}$ to generate an initial dirty image, the general form of the objective function (\autoref{eq::objfunc}) allows RML methods to fit directly to robust data products $\mathbf{d}$ derived from complex visibilities even if the visibilities themselves are corrupted by systematic errors \citep[e.g.][]{BSMEM_94, SQUEEZE, Thiebaut_2013, Katie_2015, Thiebaut_2017, Akiyama_bs, Chael18_Closure}.
High-frequency VLBI observations in particular are severely subject to rapidly varying random phase errors $\phi$ from propagation delays through atmospheric turbulence. VLBI measurements are also corrupted by station-dependent amplitude gain terms $G$ from errors in the total flux density calibration of different telescopes in the array.
In general, these station-based systematic effects combine with random thermal noise $\epsilon_{\rm ab}$ on a given baseline $a-b$ to produce a measured visibility:
\begin{equation}
\label{eq::gains}
 V_{\rm ab} = G_{\rm a}G_{\rm b}e^{\mathrm{i}(\phi_{\rm a}-\phi_{\rm b})}\,\tilde{I}(u,v) + \epsilon_{\rm ab}.
\end{equation}

Because the most severe systematic effects on the measured visibility $V_{\rm ab}$ are decomposable into station-based gains $G$ and phases $\phi$ on the separate sites $a$ and $b$, forming certain combinations of visibilities can cancel out these effects. These well-known robust `closure' products include the closure phase $\psi_{\mathcal{C},\mathrm{abc}}$ around a baseline triangle:
\begin{equation}
 \label{eq::cphase}
 \psi_{\mathcal{C},\mathrm{abc}} = \arg \left[V_{\rm ab}\,V_{\rm bc}\,V_{\rm ca}\right],
\end{equation}
and closure amplitudes $A_{\mathcal{C},\mathrm{abcd}}$ on station quadrangles:
\begin{align}
\label{eq::camps}
 A_{\mathcal{C},\mathrm{abcd}} =  \frac{|V_{\rm ab}||V_{\rm cd}|}{|V_{\rm ac}||V_{\rm bd}|}.
\end{align}
In RML imaging, we  can directly include closure phases and amplitudes in the RML objective function $J(\mathbf{I})$ by constructing the appropriate log-likelihoods. In this paper, we use the likelihood terms defined in \citet{Chael18_Closure} for the closure phases $\mathcal{L}(\mathbf{d}_{\psi_\mathcal{C}}|\mathbf{I})$ and for the logarithms of the closure amplitudes $\mathcal{L}(\mathbf{d}_{\log A_\mathcal{C}}|\mathbf{I})$.
The likelihood terms we use neglect correlations between different closure amplitudes and phases, but a judicious selection of the minimal subset of closure quantities to use can minimize these correlations; see \citet{Blackburn_2020} for details.

\subsection{Objective Function Minimization and Gradients}
To minimize the objective function for multi-frequency imaging (\autoref{eq:objfuncI})
the \texttt{eht-imaging} software uses the
Limited-Memory BFGS gradient descent algorithm \citep{LBFGS} implemented in  the
\texttt{Scipy} package \citep{SCIPY}. In the L-BFGS algorithm, it is computationally more efficient to use analytic expressions for the objective function gradient $\d J/\d \mathbf{I}$.
\citet{Chael18_Closure} provided explicit expressions for the gradients of the log-likelihood terms for EHT data ($\d \log \mathcal{L} / \d \mathbf{I}_i$) and of the various regularizer terms ($\d S / \d \mathbf{I}_i$). These derivatives are taken with respect to the image pixels at the frequency $\nu_i$ corresponding to a given dataset. To use these gradients in multi-frequency imaging, we simply need to take the derivative of the image pixels $\mathbf{I}_i$ at a given frequency with respect to our imaged quantities $[\mathbf{y}_0, \boldsymbol\alpha, \boldsymbol\beta]$, and apply the chain rule. These derivatives are simply (from \autoref{eq:loglog2}):
\begin{align}
    \frac{\d I_{i}}{\d y_0} &= I_i, \\
    \frac{\d I_i}{\d \alpha} &= I_i \, x_i, \\
    \frac{\d I_i}{\d \beta} &= I_i \, x_i^2.
\end{align}

Here, we use the same regularizer terms presented for single-frequency imaging in \citet{Chael18_Closure} on the reference image ($S_I$ in \autoref{eq:objfuncI}). The total variation and $\ell_2$ regularizer terms we use on the spectral index and spectral curvature map ($S_\alpha$ and $S_\beta$ in \autoref{eq:objfuncI}) are straightforward generalizations of corresponding regularizers used on total intensity images in \citep{Chael18_Closure}, and so the gradients of these terms are simple to adapt from those already implemented for total intensity images in \texttt{eht-imaging}.

\subsection{Imaging procedure in \texttt{eht-imaging}}
In summary, the quantities we solve for in multi-frequency RML imaging are the logarithm of the reference image $\mathbf{y}_0$, the spectral index map $\boldsymbol\alpha$ at the reference frequency $\nu_0$, and potentially the spectral curvature map $\boldsymbol\beta$ (\autoref{eq:loglog2}). In \texttt{eht-imaging}, each of these maps is defined on a fixed, square $n \times n$ grid with pixels of size $\Delta\theta$. We solve for these three images simultaneously by minimizing the objective function \autoref{eq:objfuncI}.

The image pixels in our total-intensity reconstructions are constrained to be non-negative:
\begin{equation}
    I_{i,lm} \in [0,\infty).
\end{equation}
As a result, we need to enforce some positivity constraint on the images $\mathbf{I}_i$. We employ the same strategy as in \citet{Chael18_Closure} and solve directly for the logarithm of the reference image $\mathbf{y}_0$ (\autoref{eq:transforms}) instead of $\mathbf{I}_0$ itself. Conveniently, this is already the natural parameterization in our spectral expansion in log-log space (\autoref{eq:loglog2}).

In \texttt{eht-imaging}, the pixel arrays at each frequency $\mathbf{I}_i$ parameterize a continuous function via convolution of the pixel grid with a triangular ``pulse'' function of width $2\Delta\theta$ \citep{Katie_2015}.
At each step in the minimization, we generate synthetic data from the trial images at each frequency using the Nonequispaced Fast Fourier Transform (NFFT; \citealt{NFFT}).

All of the real and synthetic datasets considered in this paper include systematic gain and phase errors at each station that vary as a function of time. In generating multi-frequency RML reconstructions, we usually begin by constructing likelihoods from the closure phases ($\psi_{\mathcal{C}}$) and log-closure amplitudes ($\log A_\mathcal{C}$); we fit directly to these robust data products through \autoref{eq:objfuncI}. After generating images and spectral index/curvature maps from the closure quantities, however, it is often possible to generate more accurate reconstructions by self-calibrating the data and re-imaging. The self-calibration (or ``hybrid mapping'') procedure involves solving for the gain terms $G$ and phase terms $\phi$ in \autoref{eq::gains} given a reconstruction $\mathbf{I}$ of the on-sky image $I(x,y)$. Self-calibration is useful for RML imaging, but essential for CLEAN imaging of datasets with phase or amplitude errors, as CLEAN requires calibrated data to generate an initial dirty image for deconvolution \citep[e.g.,][]{Wilkinson77,Readhead78,Readhead80,Schwab_1980, Cornwell_1981,Pearson_1984,Walker_1995,Cornwell_1999}.

In the RML imaging in this paper, we usually self-calibrate only once to an initial image generated by fitting closure phases and amplitudes. We then re-image by fitting to the calibrated visibility amplitudes $|V|$ and closure phases $\psi_\mathcal{C}$. We assume that we have good independent measurements of the total flux density and re-scale the closure-only image results (which are insensitive to total flux) to the correct values before self-calibrating.
Since small errors in the phase calibration can easily lead to large image artifacts, and since for reasonably large arrays the phase information contained in closure phases is a significant fraction of the information in the full complex visibility phases \citep{TMS}, we do not fit directly to complex visibilities for most of the reconstructions presented here (with one exception in \autoref{sec:realdata}).

To avoid local minima in the objective function, we follow the procedure introduced in \citet{Chael18_Closure} and run each objective function optimization multiple times with different initial conditions. After finishing one minimization round (from reaching convergence on the gradient of $J$ or reaching the maximum number of allowed gradient descent steps), we restart the optimization from a blurred version of the output image. This procedure aids in convergence to a global minimum and helps the imager remove spurious high-frequency artifacts that are unconstrained by the data.

\section{Example Reconstructions: Synthetic ngEHT data}
\label{sec:ngEHT}

\begin{figure*}[t]
\centering
\includegraphics[width=\textwidth]{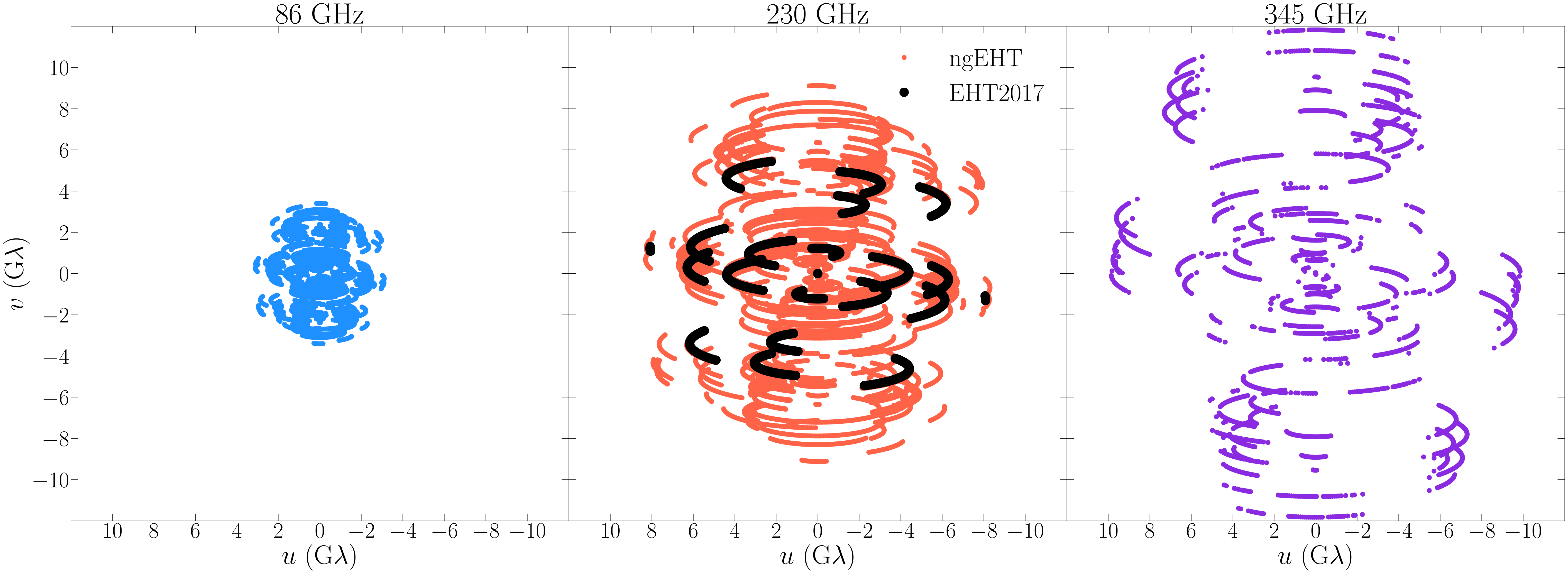}
\caption{$u-v$ coverage at 86 GHz (left) 230 GHz (center) and 345 GHz (right) for the ngEHT concept array used in this work. The center panel shows the 2017 EHT coverage in black, for reference.}
\label{fig:ngEHTuv}
\end{figure*}

\begin{figure*}[t]
\centering
\includegraphics[width=\textwidth]{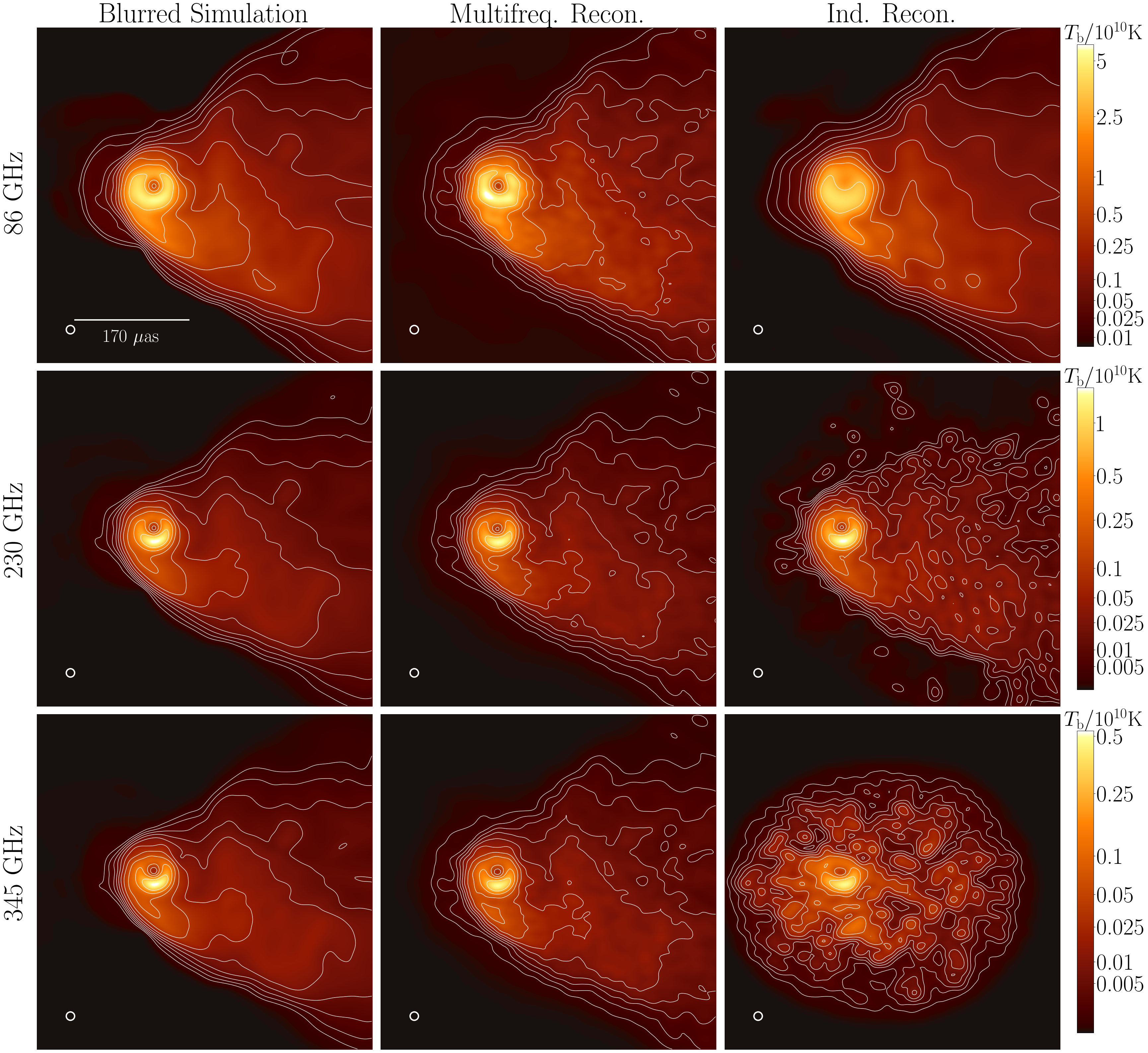}
\caption{Image reconstructions of a radiative MAD GRMHD model of M87* from \citet{ChaelM87}. Rows from top to bottom show the ground truth images and ngEHT reconstructions at 86, 230, and 345 GHz. The leftmost column shows the ground truth simulation images blurred with a circular Gaussian kernel with a FWHM of 12$\mu$as, 1/2 of the ngEHT nominal resolution at 230 GHz. The second column shows reconstructions of these images from ngEHT data using multi-frequency synthesis, combining information across the three frequency bands. The last column shows images reconstructed independently at each frequency band from the same data as the images in the second column. Contours indicate successive powers of $1/2$ from the peak brightness point in each image.}
\label{fig:rjetI}
\end{figure*}

\begin{figure*}[t]
\centering
\includegraphics[width=\textwidth]{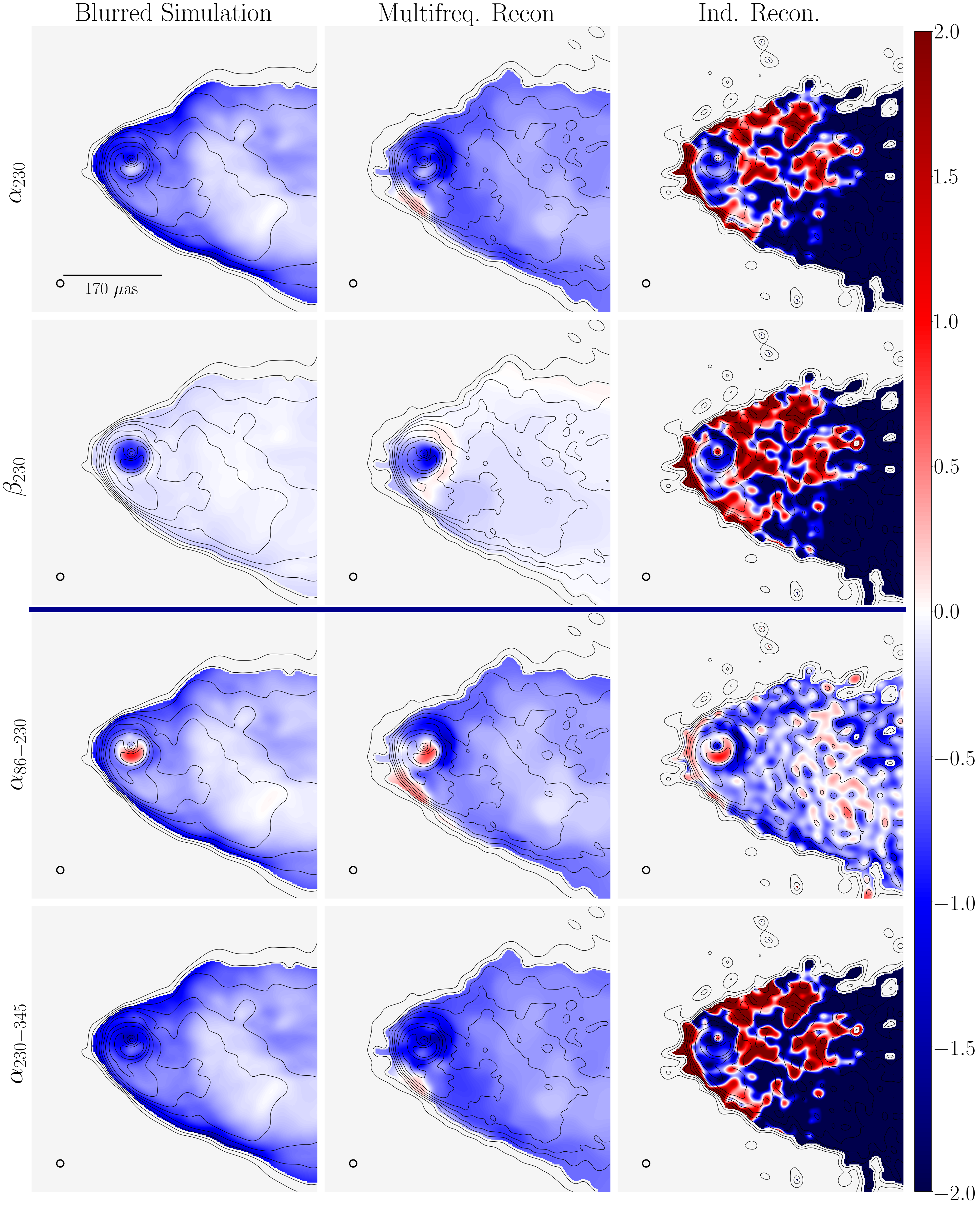}
\caption{Spectral index and curvature maps from reconstructions of the radiative MAD GRMHD model in \autoref{fig:rjetI}. As in \autoref{fig:rjetI}, the columns from left to right correspond to the blurred ground truth simulations, the multi-frequency synthesis reconstructions, and the reconstructions done independently at the three frequencies. The top row shows the spectral index at 230 GHz. The spectral index is fit directly to data in the multi-frequency reconstruction; we compute the corresponding index for the single-frequency reconstructions by fitting a second-order polynomial in $\log\nu$ and $\log I_\nu$ for each pixel after imaging. The second row shows the spectral curvature at 230 GHz obtained in the same way. The third row shows the spectral slope between the 86 GHz and 230 GHz images (derived from fitting a line in $\log\nu$ and $\log I_\nu$ from just those two images) and the final row shows the spectral slope between 230 and 345 GHz derived in the same way.
}
\label{fig:rjetspec}
\end{figure*}

\begin{figure*}[t]
\centering
\includegraphics[width=\textwidth]{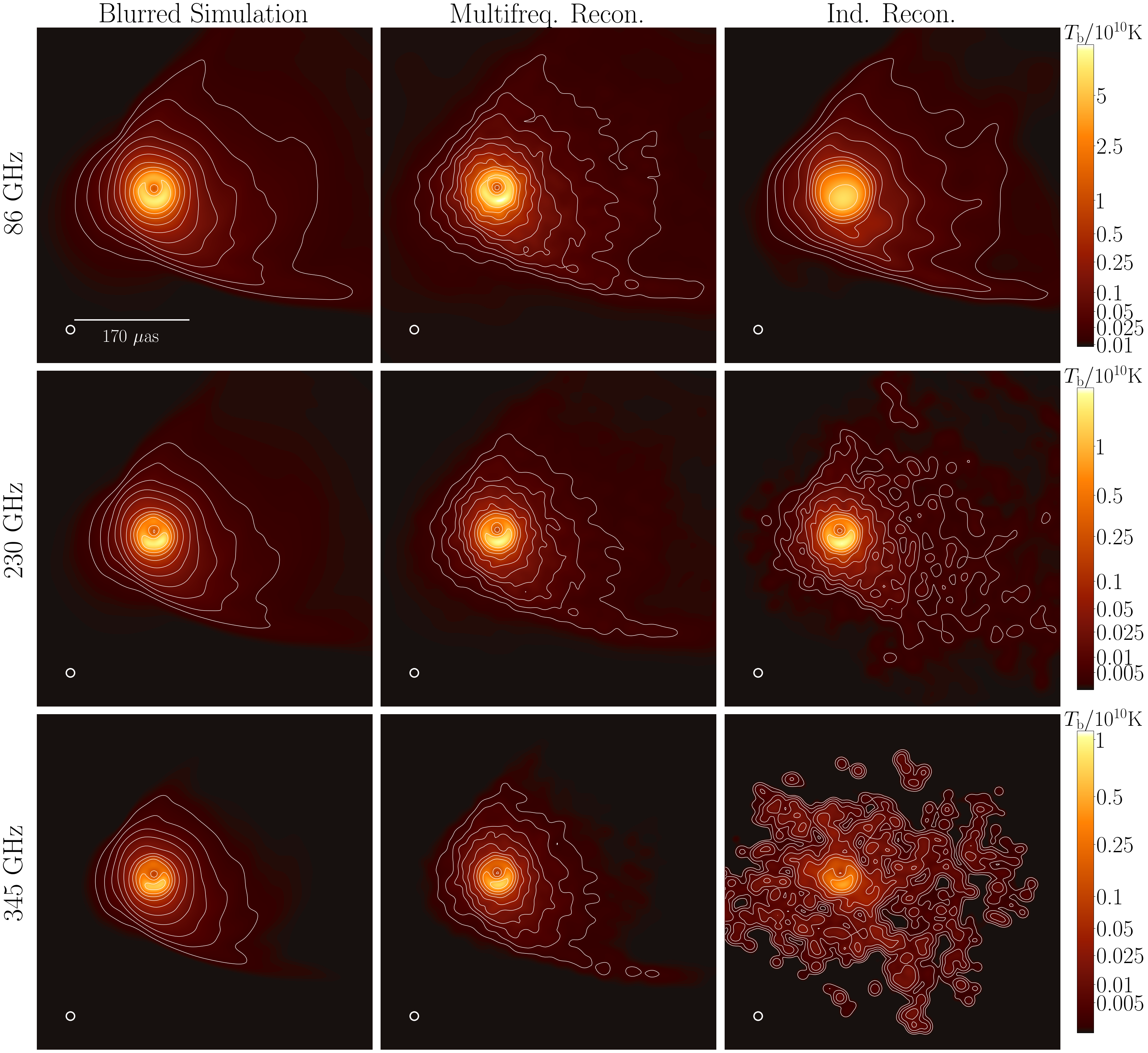}
\caption{Same as \autoref{fig:rjetI}, but for an M87* jet model from \citet{Mizuno2021} averaged across 2000\,M.}
\label{fig:cjetI}
\end{figure*}

\begin{figure*}[t]
\centering
\includegraphics[width=\textwidth]{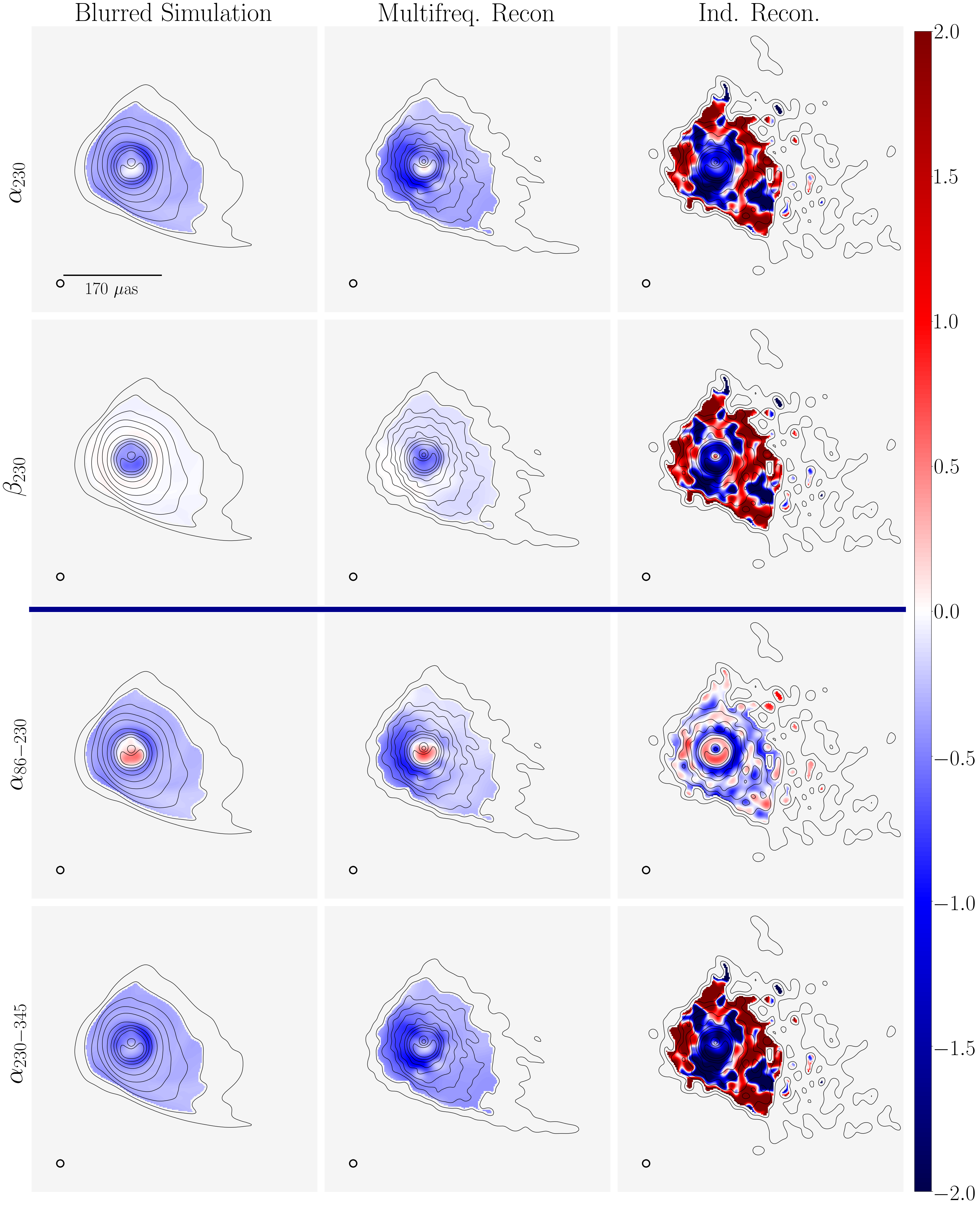}
\caption{Same as \autoref{fig:rjetspec}, but for the M87* jet model from \citet{Mizuno2021}.}
\label{fig:cjetspec}
\end{figure*}

\begin{figure*}[t]
\centering
\includegraphics[width=.9\columnwidth]{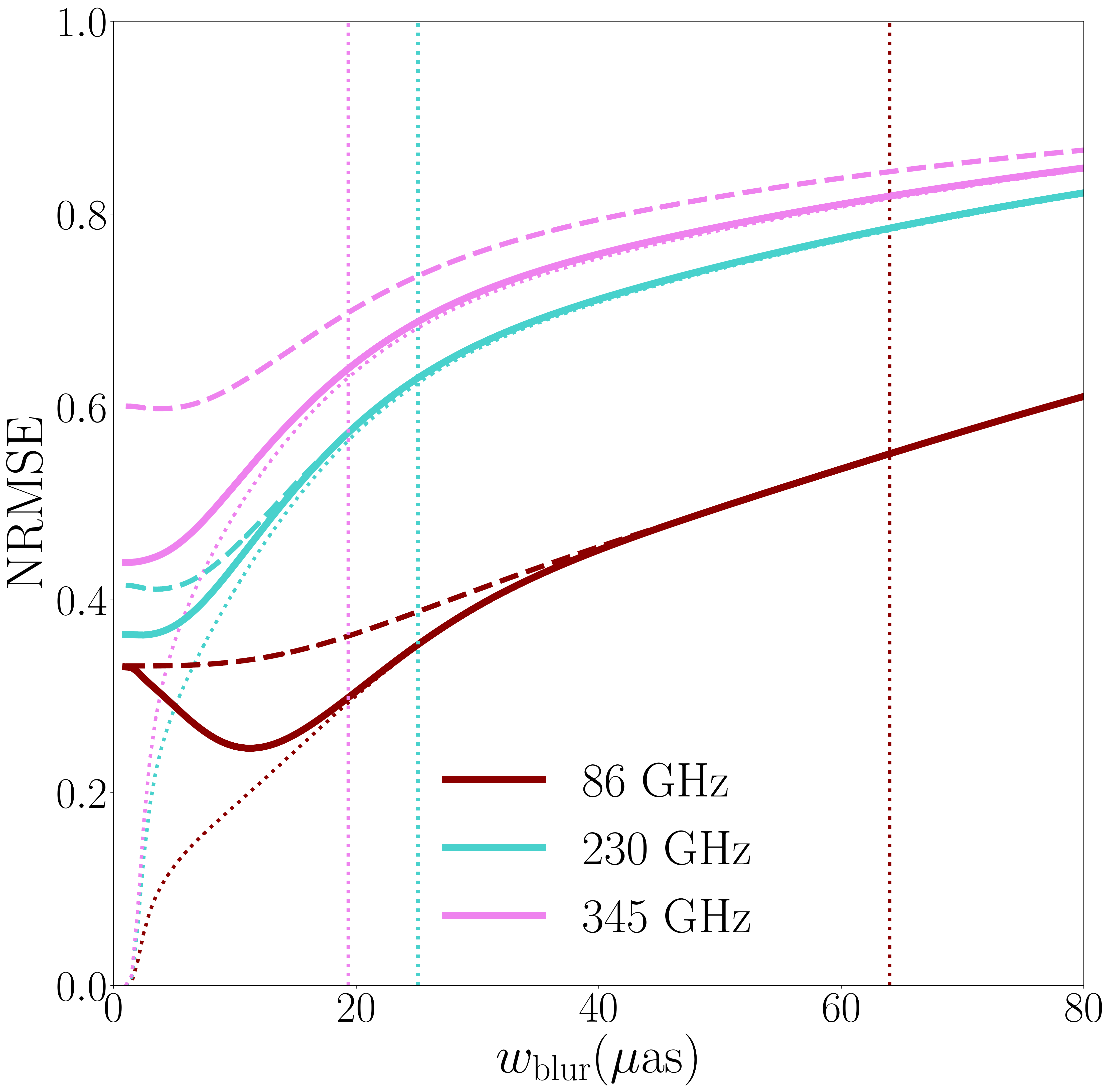}
\includegraphics[width=.9\columnwidth]{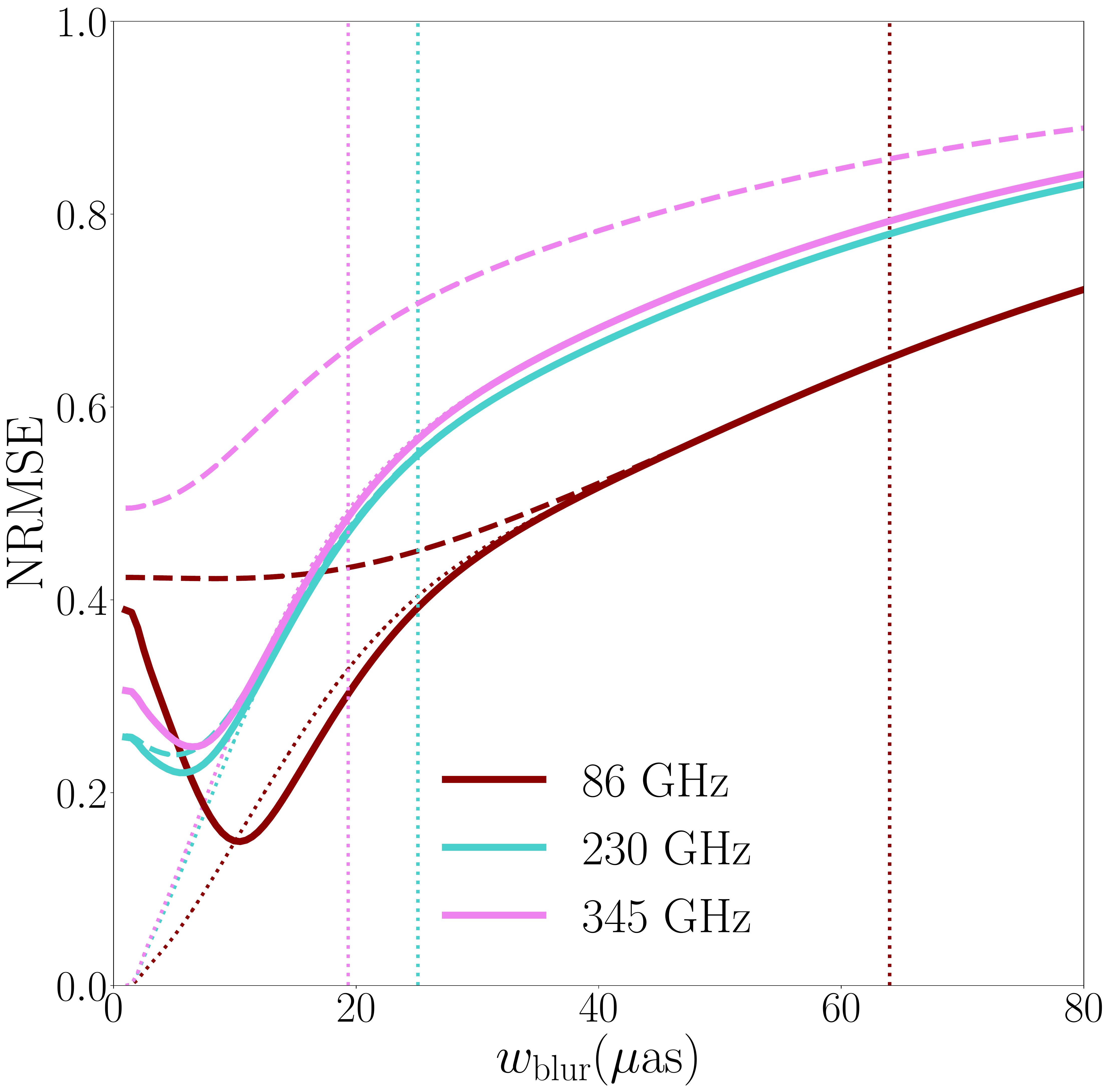}
\caption{(Left) Curves of the normalized root-mean-square error (NRMSE, \autoref{eq::nrmse}) image fidelity metric with restoring beam size for the ngEHT multi-frequency reconstructions (solid lines) and independent single-frequency reconstructions (dashed lines) at 86, 230, and 345 GHz of the model shown in \autoref{fig:rjetI}.
We compute the NRMSE by comparing an image reconstruction blurred with a Gaussian kernel of FWHM $w_{\rm blur}$ with the unblurred ground truth image. The dotted lines indicate the NRMSE obtained by comparing the ground truth image with the blurred version of itself. Vertical lines indicate the nominal resolution (1/$|u|_{\rm max}$) of the ngEHT array at each frequency. (Right) The same as in the left panel, but for the model shown in \autoref{fig:cjetI}.
}
\label{fig:superres}
\end{figure*}

In this section, we present multi-frequency reconstructions  of synthetic GRMHD simulation models of the emission from M87* as observed by the ngEHT. For ngEHT station locations and telescope parameters, we use the proposed ngEHT reference array presented in \citet{Issaoun_2022} that was developed from the site-selection considerations presented in \citet{Raymond_2021}. The ngEHT sites we use in generating synthetic data are summarized in \autoref{tab:ngEHTarray} and are mapped in \autoref{fig:ngEHT}.

In \autoref{sec:syndata} we describe our procedure for generating synthetic ngEHT data. In \autoref{sec:ngEHT86-345} we present reconstructions of two GRMHD simulation data sets and their spectral index and curvature maps from observations taken across the full proposed ngEHT frequency range at 86, 230, and 345 GHz. In \autoref{sec:ngEHT230}, we present reconstructions of two different GRMHD data sets and their spectral index maps from four ngEHT spectral windows clustered near 230 GHz.\footnote{The synthetic data, ground truth simulation images, and \texttt{eht-imaging} scripts used to produce the reconstructed images in the following sections can be found at https://github.com/achael/multifrequency\_scripts/.}

\subsection{Generating synthetic ngEHT data}
\label{sec:syndata}

In the examples that follow in \autoref{sec:ngEHT}, we use the \texttt{eht-imaging}  library to generate synthetic data on ngEHT baselines. The \texttt{eht-imaging} synthetic data generation code generates \uv  coverage for an interferometric array at a given observing frequency, samples the Fourier transform on the appropriate baselines, and adds random thermal noise and systematic gain and phase errors following \autoref{eq::gains}.\footnote{\texttt{eht-imaging}'s synthetic data generation can also add polarimetric leakage errors and right- and left-circular polarization gain offsets, both of which are ignored in this work.} The thermal noise and gain terms in \texttt{eht-imaging} are generated from probability distributions defined by the user. Sophisticated end-to-end VLBI synthetic data generation tools like \texttt{SYMBA} \citep{Roelofs_2020} can also be used to generate errors from full modeling of telescope and atmosphere parameters; given suitable parameter choices, \texttt{eht-imaging} and \texttt{SYMBA} have been shown to produce equivalent datasets for imaging purposes \citep{PaperIV}.

The thermal noise on a given baseline between site $a$ and site $b$ is
\begin{equation}
\label{eq::noise}
\sigma_{\rm ab} = \frac{1}{\eta} \sqrt{\frac{\text{SEFD}_\mathrm{a} \times\ \text{SEFD}_\mathrm{b}}{2\, \Delta \nu \, \Delta t}},
\end{equation}
where  $\Delta\nu$ is the bandwidth, $\Delta t$ is the integration time, and $\eta$ is an efficiency factor relating to the data quantization (for 2-bit quantization used by the EHT, $\eta=0.88$). In the ngEHT examples in this section, we assume a bandwidth $\Delta\nu=2$GHz for each ngEHT sub-band and we use an integration time $\Delta t=120$s. The factors $\text{SEFD}_{\rm a}$ and $\text{SEFD}_{\rm b}$ in \autoref{eq::noise} are the ``system equivalent flux densities'' of the two telescopes in the baseline pair.

In synthetic data generation for this paper, we determine SEFDs for each station as a function of time (rather than assuming a single fixed value) by following a similar procedure to \cite{Raymond_2021}.  The effective SEFD at each station is given by
\begin{equation}
\text{SEFD} = \frac{2k_{\rm B}T_\text{sys}}{A_\text{eff}} e^{\tau} ,
    \label{eq::SEFD}
\end{equation}
\noindent where $\tau$ is the line-of-sight atmospheric opacity at the observing frequency, $A_\text{eff}$ is the effective collecting area of the telescope,
\begin{equation}
T_{\text{sys}} = T_{\text{rx}} + T_{\text{atm}} \left( 1 - e^{-\tau} \right) \label{eq::Tsys}
\end{equation}
\noindent is the system temperature, and $T_{\text{atm}}$ is the temperature of the atmosphere.  We assume receiver temperatures $T_{\text{rx}}$ of 40\,K at 86\,GHz, 50\,K at 230\,GHz, and 75\,K at 345\,GHz.  We use historical atmospheric data from the MERRA-2 database \citep{Gelaro_2017} as inputs to the \textit{am} radiative transfer code \citep{Paine_2019}, which then provides atmospheric opacity $\tau$ and atmospheric temperature $T_{\text{atm}}$ at each site as a function of observing frequency. These quantities, along with the collecting area and aperture efficiency of each telescope, determine the level of source attenuation and station SEFD as a function of time through \autoref{eq::SEFD}.

The impact of atmospheric effects on the synthetic data is most pronounced for 345\,GHz observations, where it is typical for $\sim$50\% of the array to be operating with $\tau > 1$.  The resulting decrease in sensitivity limits the overall number of detected data points and reduces the signal-to-noise ratio of the data points that are detected, relative to observations at lower frequencies.  We note that for the ngEHT, it may be possible to use frequency phase transfer (FPT) techniques \citep[e.g.,][]{Rioja_2011,Dodson_2017,Rioja_2020} to bootstrap atmospheric phase solutions from lower-frequency observations up to 345 GHz, thereby substantially extending the coherent integration time and enabling additional detections\citep[e.g.,][]{Issaoun_2022}.  However, in this paper we do not simulate observations that take advantage of FPT techniques.

\subsection{ngEHT Case Study: Simultaneous imaging from 86-345 GHz}
\label{sec:ngEHT86-345}

In this section we consider two examples of ngEHT data reconstruction from synthetic observations at 86, 230, and 345 GHz.\footnote{The synthetic data, ground truth simulation images, and \texttt{eht-imaging} scripts used to produce the results in the following Sections can be found at https://github.com/achael/multifrequency\_scripts/.} We produce synthetic ngEHT data from two sets of GRMHD simulation images of M87*; we then reconstruct images from the multi-frequency data using both standard RML optimization at each frequency independently and using the multi-frequency approach introduced in this paper. We compare the simultaneously-recovered maps of the reference frequency $\mathbf{I}_0$, spectral index $\boldsymbol\alpha$, and spectral curvature $\boldsymbol\beta$ from multifrequency imaging to those we compute after-the-fact from independently-reconstructed images. The \uv coverage of the synthetic ngEHT observations of M87* we use for both source models in this section is shown in \autoref{fig:ngEHTuv}.

\subsubsection{\citealt{ChaelM87} model}

\begin{table*}[ht]
    \centering
    \begin{tabular}{c|cc|ccc}
    \hline
      Source  & Method & Frequency  & $\chi^2_{\mathrm{log} A_C}$ & $\chi^2_{\psi_C}$ & $\chi^2_{|V|}$  \\
      \hline \hline
      \citet{ChaelM87} M87 Jet (\autoref{fig:rjetI}) & Single-Frequency & 86 GHz & 1.04 & 1.00 & 0.84  \\
       & & 230 GHz & 0.98  & 1.00 & 0.84 \\
       & & 345 GHz & 1.11 & 1.20 & 0.71  \\
       & Multi-Frequency & 86 GHz & 1.07 & 1.03 & 0.88 \\
       & & 230 GHz & 1.02 & 1.04 & 0.88 \\
       & & 345 GHz & 1.14 & 1.23 & 0.69 \\
      \citet{Mizuno_2021} M87 Jet (\autoref{fig:cjetI}) & Single-Frequency & 86 GHz & 1.06 & 1.03 & 0.86 \\
       & & 230 GHz & 0.98 & 0.99 & 0.82 \\
       & & 345 GHz & 1.30 & 1.17 & 0.94 \\
       & Multi-Frequency & 86 GHz & 1.09 & 1.07 & 0.90 \\
       & & 230 GHz & 1.03 & 1.02 & 0.88  \\
       & & 345 GHz & 1.34 & 1.27 & 0.86 \\
       \hline
      \citet{Ricarte_2022} MAD (\autoref{fig:angeloMAD}) & Single-Frequency & 213 GHz & 0.98 & 1.09 & 0.87 \\
       & & 215 GHz & 0.97  & 1.09 & 0.85 \\
       & & 227 GHz & 1.01 & 1.11 & 0.87  \\
       & & 229 GHz & 0.96 & 1.11 & 0.86 \\
       & Multi-Frequency & 213 GHz & 0.98 & 1.09 & 0.84 \\
       & & 215 GHz &  0.98 &  1.10 & 0.84 \\
       & & 227 GHz &  1.02 & 1.11 & 0.84  \\
       & & 229 GHz & 0.97 & 1.11 & 0.84 \\
      \citet{Ricarte_2022} SANE (\autoref{fig:angeloSANE}) & Single-Frequency & 213 GHz & 1.15 & 1.11 & 0.97 \\
       & & 215 GHz & 1.04 & 1.11 & 0.93  \\
       & & 227 GHz & 1.03 & 1.11 & 0.94  \\
       & & 229 GHz & 1.04 & 1.11  & 0.97 \\
       & Multi-Frequency & 213 GHz & 1.18 & 1.12 & 0.89 \\
       & & 215 GHz & 1.07 & 1.12 & 0.89  \\
       & & 227 GHz & 1.04 & 1.12 & 0.88  \\
       & & 229 GHz & 1.04 & 1.12  & 0.88\\
       \hline
    \end{tabular}
    \caption{Reduced $\chi^2$ table for the simulated data sets in \autoref{sec:ngEHT}. We provide reduced-$\chi^2$ statistics comparing the final self-calibrated data sets for each frequency band with the final image reconstructions for the data products used in the fit: log closure amplitudes ($\chi^2_{\mathrm{log}A_{\mathcal{C}}}$), closure phases ($\chi^2_{\psi_{\mathcal{C}}}$), and visibility amplitudes ($\chi^2_{|V|}$).}
    \label{tab:chisqs_snyth}
\end{table*}

We first consider synthetic images of the near-horizon accretion flow and jet in M87* from a radiative GRMHD simulation (simulation \texttt{R17} from \citealt{ChaelM87}). The simulation was performed using the radiative GRMHD code \texttt{KORAL} \citep{KORAL13,KORAL14,KORAL16}.
The simulation is in the magnetically arrested (MAD)  state of black hole accretion \citep{Igumenschchev2003,NarayanMAD}, which is favored by polarimetric EHT observations of M87* \citep{PaperVIII}.
The black hole spin was set to $a_{\star}=0.9375$. The electron distribution was evolved self-consistently in the simulation under synchrotron cooling and sub-grid heating from magnetic reconnection, using results from \citet{Rowan17}. We generated
images of the 345~GHz, 230~GHz, 86~GHz synchrotron emission from the simulation using the GR  radiative transfer code \texttt{grtrans} \citep{Dexter16}.

We present total intensity reconstructions of the simulated ngEHT data taken from these simulation images in \autoref{fig:rjetI} and we show spectral index reconstructions in \autoref{fig:rjetspec}. In \autoref{fig:rjetI}, we compare the simulation images at each frequency blurred with a $12\mu$as circular Gaussian kernel (left column) to \texttt{eht-imaging} reconstructions performed using multi-frequency synthesis (center) and reconstructions performed independently at each frequency (right column). The data term hyperparameters $\kappa$ and total intensity regularizer hyperparameters $\lambda_I$ were fixed to the same values in all cases, both for each independent single-frequency imaging script as well as in the multi-frequency optimization.

\autoref{fig:rjetI} shows that the ngEHT coverage is sufficiently dense and its sensitivity sufficiently high to recover good independent reconstructions of structure in M87*'s core and extended jet at both 86 and 230 GHz. At 345 GHz, the ngEHT's sensitivity is much lower; as a result, the image reconstructed independently using 345 GHz data alone recovers the central ring structure but does not recover the extended low-brightness emission from the jet.\footnote{The size of the distribution of low-brightness noisy structure in the single-frequency 345 GHz reconstructions is largely determined by the size of the initial Gaussian model used to start the RML optimization.} However, when we image all three frequencies simultaneously, information is shared across the three frequencies and structural information from the 86 GHz and 230 GHz observations can serve as an effective ``regularizer'' on the 345 GHz reconstruction. As a result, the reconstruction of the 345 GHz extended jet emission is much more accurate at high dynamic range in the multi-frequency reconstruction.

Furthermore, while the independent reconstructions at all three frequencies show evidence for superresolution of structures smaller than the EHT nominal resolution, this super-resolving power is enhanced at the lower frequencies in the multi-frequency reconstruction. The 86 GHz simulation image has an optically thick core but optically thin jet; as a result, a central brightness depression is visible at 86 GHz. In this simulation, the 86 GHz central brightness depression is closely associated with the black hole's `inner shadow', or the lensed image of the event horizon's boundary in the equatorial plane \citep[see][Figure 6]{Chael_21}.
This brightness depression is not clearly resolved by the independent image reconstruction at 86 GHz. However, the multi-frequency reconstruction accurately resolves the inner shadow feature at 86 GHz by propagating structural information from the higher-resolution datasets at 230 and 345 GHz.

\autoref{fig:rjetspec} shows the performance of the two methods (independent single-frequency RML imaging and multi-frequency RML imaging) at recovering resolved spectral index information. Because spectral curvature is significant, the multi-frequency method applied here directly reconstructs both the spectral index map at the 230 GHz reference frequency ($\boldsymbol\alpha$, top row of \autoref{fig:rjetspec}) and the spectral curvature ($\boldsymbol\beta$, second row of \autoref{fig:rjetspec}). In the bottom two rows, we show maps of the spectral slope computed between the two pairs of frequencies: 86-230 GHz and 230-345 GHz. In computing spectral index and curvature maps from the images independently reconstructed at each frequency (right column of \autoref{fig:rjetspec}), we first align the images to each other by finding the image shift that maximizes their cross-correlation.

Independent single-frequency ngEHT imaging can accurately recover some details of the spectral index structure between 86-230 GHz (right column, third row of \autoref{fig:rjetspec}). The independent reconstructions at these frequencies accurately reconstruct the positive spectral index from optically thick material in the accretion disk and the flat or slightly negative spectral index of optically thin material in the jet. However, the images independently reconstructed at the three frequencies do not well constrain the spectral slope between 230-345 GHz (fourth row) or the spectral curvature map between all three frequencies (second row). By contrast, the multi-frequency RML reconstruction produces good reconstructions of both the spectral index $\boldsymbol\alpha$ and curvature $\boldsymbol\beta$ across most of the image at the reference frequency of 230 GHz. As a result, multi-frequency RML synthesis can recover an accurate map of the spectral index between 230-345 GHz (fourth row), and the recovered structure in the 86-230 GHz spectral index map (third row) is more accurate than in the independent reconstructions, particularly in the extended, low-brightness jet.

The spectral index recovery from multi-frequency RML synthesis is not without artifacts; in particular, an anomalously high spectral index on the bottom edge of the jet close to the central black hole is apparent in \autoref{fig:rjetspec}. This error in the spectral index reconstruction occurs at the jet edge where the brightness gradient is very steep. In the Appendix, we show that this artifact can be mitigated by increasing the hyperparameter values $\lambda_{TV}$ on the spectral index total variation regularizer term in the objective function (\autoref{eq:alphatv}). In RML imaging in general, it is advisable to survey over the space of hyperparameters to determine which values are best suited for a particular data set \citep{PaperIV}.  Here, we present the reconstruction in \autoref{fig:rjetspec} as our fiducial result both because it uses moderate values of the total variation hyperparameter that we found worked reasonably well over a large range of synthetic data sets, and because it presents a cautionary tale that even improved methods for image or spectral image recovery from sparse VLBI datasets are not immune to artifacts.

In \autoref{tab:chisqs_snyth}, we provide summary reduced-$\chi^2$ statistics to quantify the goodness-of-fit to the data in the single-frequency and multi-frequency image reconstructions in \autoref{fig:rjetI}. We present $\chi^2$ values comparing the reconstructed images to the final self-calibrated synthetic data set  for the three quantities we fit to: the log closure amplitudes $\chi^2_{\mathrm{log}A_\mathcal{C}}$, the closure phases $\chi^2_{\psi_\mathcal{C}}$, and the visibility amplitudes $\chi^2_{|V|}$.\footnote{The exact definitions of the reduced $\chi^2$ statistics we report can be found in \citealt{Chael18_Closure} equations 21, 19, and 17 for log closure amplitudes, closure phases, and visibility amplitudes, respectively.} For good image reconstructions, we expect these $\chi^2$ terms to be close to unity. However, since the definitions of the reduced $\chi^2$ terms we use ignore correlations between closure phase and amplitude measurements and the non-Gaussianity of these quantities at low SNR, these values should be interpreted with care \citep{Blackburn_2020}. Furthermore, the visibility amplitudes $|V|$ are adjusted by self-calibrating the station gains to intermediate image results, so we expect the values of $\chi^2_{|V|}$ reported to be systematically below unity. The results in \autoref{tab:chisqs_snyth} indicate that all the image reconstructions in \autoref{fig:rjetI} provide satisfactory fits to the data. Furthermore, there is no clear difference in the $\chi^2$ statistics between the single- and multi-frequency reconstructions; both fit the data well, and differences between the two reconstructions arise in differences in the underlying image model and regularizing terms used in the imaging process.

\subsubsection{\citealt{Mizuno2021} model}
Figures \ref{fig:cjetI} and \ref{fig:cjetspec} show total intensity and spectral index reconstructions of another GRMHD model of M87* from \citet{Mizuno2021}. The simulation images display the time-averaged structure across 2000 gravitational times $t_{\rm g}$ ($\sim$ 2 years for M87*).
The simulations used the GRMHD code \texttt{BHAC}
\citep{Porth2017, Olivares2020} using three layers of adaptive mesh refinement in logarithmic Kerr-Schild coordinates.
The simulation assumes a black hole spin $a_{\star}=0.9375$
The simulation was evolved to 30000 $t_{\rm g}$; during this time period the simulation is well within the MAD state.
The radiative transfer calculations to produce the images in \autoref{fig:cjetI} used the GRRT code \texttt{BHOSS}  \citep{Younsi2012,Younsi2020}. The radiative post-processing
assumed a kappa distribution for the emitting electrons in the jet following \citet{Davelaar2019} and \citet{Fromm2021}, using the results of to determine the power-law index \citet{Ball2018}.
In order to match the simulation images to observations we adjusted the simulation mass accretion rate to produce an average flux of $\approx1.0$\,Jy at 230\,GHz over a time window from 28000-30000\,$t_{\rm g}$.

The assumed ngEHT baseline coverage, sensitivity, and \texttt{eht-imaging} reconstruction scripts for this model were identical to those used in the reconstructions of the \citet{ChaelM87} model in Figures \ref{fig:rjetI} and \ref{fig:rjetspec}. The ngEHT reconstructions of the \citet{Mizuno_2021} model in Figures \ref{fig:cjetI} and \ref{fig:cjetspec} show similar features to those seen in the earlier example. While independent ngEHT imaging can recover details of the core structure and extended jet emission at 86 and 230 GHz, lower sensitivity at 345 GHz makes it difficult to extract information on the extended jet from this frequency alone. When combining data from all three frequencies in multi-frequency synthesis, the 345 GHz reconstruction accurately reproduces the extended jet structure and the 86 GHz reconstruction super-resolves the central brightness depression. The multi-frequency reconstruction accurately recovers both the spectral index map from 86-230~GHz (weakly recovered by independent imaging) and from 230-345 GHz (not recovered by independent imaging). In \autoref{tab:chisqs_snyth}, we present the reduced-$\chi^2$ statistics quantifying the fit quality for this reconstruction.

\subsubsection{Quantifying Reconstruction Fidelity}
We can quantify the fidelity of the reconstructions of the two models presented in \autoref{fig:rjetI} and \autoref{fig:cjetI} as a function of restoring beam size \citep[e.g.][Figure 4]{Chael_16}. For both GRMHD models, we take the ngEHT multi-frequency reconstruction at each observed frequency (86, 230, and 345 GHz) and blur the results with circular Gaussian beams of increasing FWHM size $w_{\rm blur}$ between 0 and 80 $\mu$as. For each beam-convolved reconstruction image, we compute the normalized root-mean square error (NRMSE) with the unblurred simulation image (the `ground-truth') at the appropriate frequency. The NRMSE of the blurred reconstruction $\mathbf{R}$ compared to the ground truth $\mathbf{G}$ is
\begin{equation}
\label{eq::nrmse}
 \text{NRMSE}(\mathbf{R},\mathbf{G}) = \frac{\sqrt{\sum_{lm}(R_{lm} - G_{lm})^2}}{{\sqrt{\sum_{lm}{G_{lm}^2}}}},
\end{equation}
where the sums are over all pixels $l,m$ in both dimensions.
Before computing the NRMSE, we align the images in order to maximize the normalized cross-correlation between the ground truth and the reconstruction.\footnote{We use the NRMSE (\autoref{eq::nrmse}) instead of the normalized image cross-correlation (e.g. Equation 15, \citealt{PaperIV})  for consistency with the plots in \citet{Chael_16} and because NRMSE is sensitive to total flux offsets between the reconstruction and ground truth image.}
We plot the resulting NRMSE values as a function of the restoring beam FWHM size $w_{\rm blur}$ in \autoref{fig:superres} for both the single-frequency reconstructions (dashed lines) and the multi-frequency reconstructions (solid lines). For reference, we also plot the NRMSE of the blurred simulation image compared against the unblurred ground truth (dotted lines); the NRMSE of this self-comparison drops to zero at small beam sizes, while the reconstructions all have non-zero NRMSE at all beam sizes.
For both source models and for all three frequencies, we first observe that the single-frequency reconstructions all have larger NRMSE values than the corresponding multi-frequency results at all values of the blurring kernel FWHM $w_{\rm blur}$. This result indicates that the multifrequency reconstruction provides a more accurate reproduction of the original image than single-frequency imaging in both models at all three frequencies.

We can look at the beam size $w_{\rm blur}$ of the minimum NRMSE in \autoref{fig:superres} as an indication of an optimal blurring kernel size for the given reconstructions. As discussed in \citet{Chael_16}, RML images may not require a post-hoc blurring step if the regularizers are tuned to suppress spurious high-frequency structure. In this case,  the NRMSE curve as a function of $w_{\rm blur}$ will flatten to its minimum value as $w_{\rm blur}\rightarrow 0$. In some cases, RML imaging does produce spurious structure at very high spatial frequencies not sampled by the interferometer; in these cases the NRMSE curve has a prominent minimum at nonzero $w_{\rm blur}$ and increases at small blurring kernel sizes as $w_{\rm blur}\rightarrow 0$. The 86 GHz multifrequency reconstructions in \autoref{fig:superres} fall in this category.

In \autoref{fig:superres},
the NRMSE curves of all methods have minima at values smaller than the nominal resolution at that frequency:  $w_{\rm beam} < w_{\rm nominal}$. If the underlying reconstruction has structure on small spatial scales, the position of the minimum NRMSE can indicate that the RML imaging methods are superresolving source structure \citep[see also][ Figures 4 and 5]{Chael_16}. However, while the position of the NRMSE minima provide an indication of the optimal restoring beam size for a given reconstruction, it is not possible to directly equate this beam size to a superresolving scale. For example, the 86 GHz single frequency image reconstructions in the left panel of  \autoref{fig:superres} do not contain significant structure on scales smaller than $\approx20\mu$as; as a result, they have comparable NRMSE values for values of $w_{\rm blur}$ between 0 and $20\mu$as without a pronounced minimum. Nonetheless, all the reconstructions in \autoref{fig:rjetI} and \autoref{fig:cjetI} do contain features on scales smaller than the nominal beam at a given frequency and these features are not penalized by the NRMSE metric in \autoref{fig:superres}. Thus, while it remains difficult to precisely quantify the achievable level of superresolution for a given reconstruction method, taken together, \autoref{fig:rjetI}, \autoref{fig:cjetI}, and \autoref{fig:superres} indicate that both single- and multi-frequency ngEHT reconstructions are capable of super-resolving source structure, and that the multi-frequency method generally produces higher-fidelity reconstructions of the underlying source.


\subsection{ngEHT Case Study: 230 GHz spectral index maps across the band}
\label{sec:ngEHT230}

\begin{figure*}[t]
\centering
\includegraphics[width=.9\columnwidth]{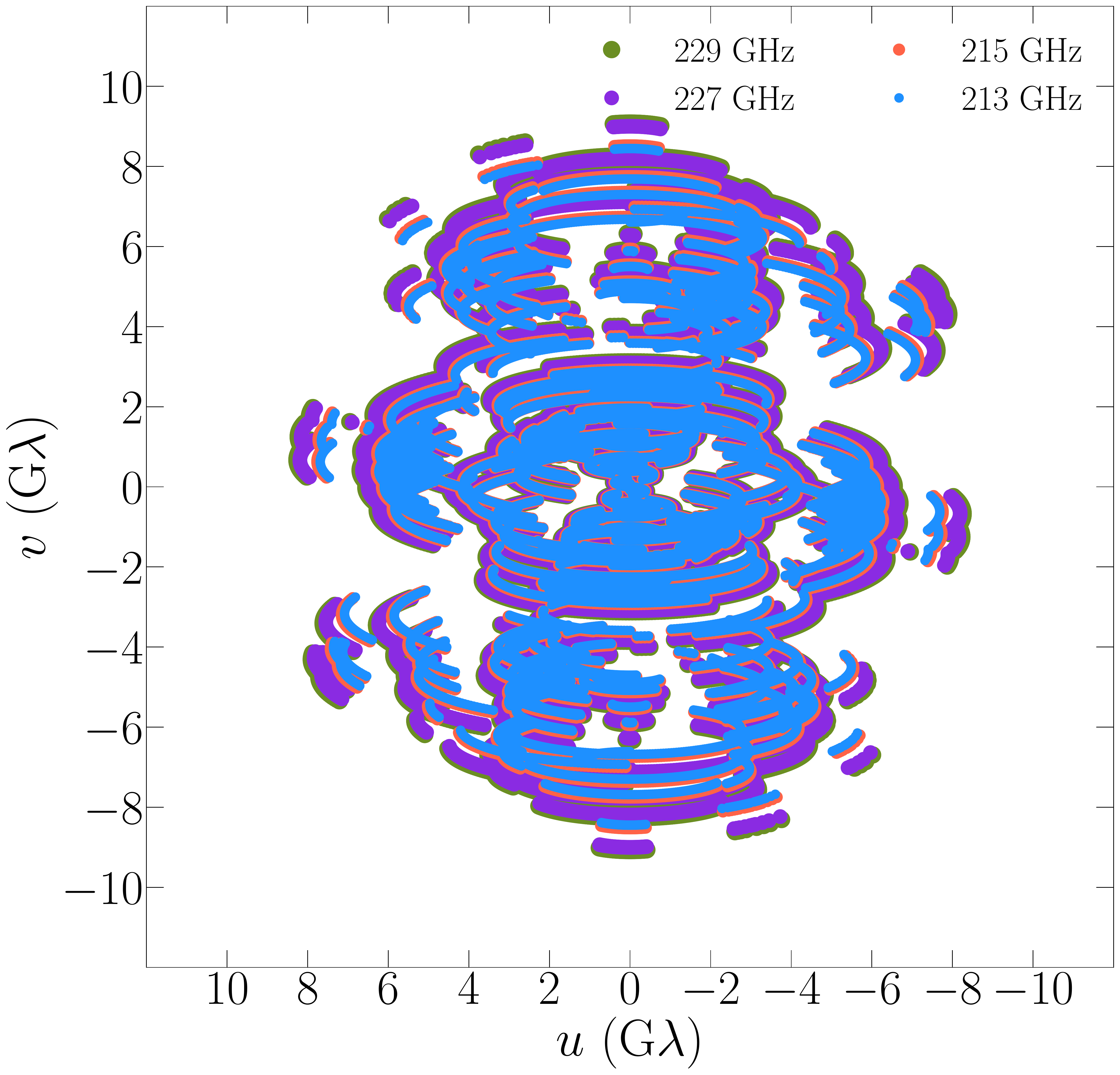}
\includegraphics[width=.9\columnwidth]{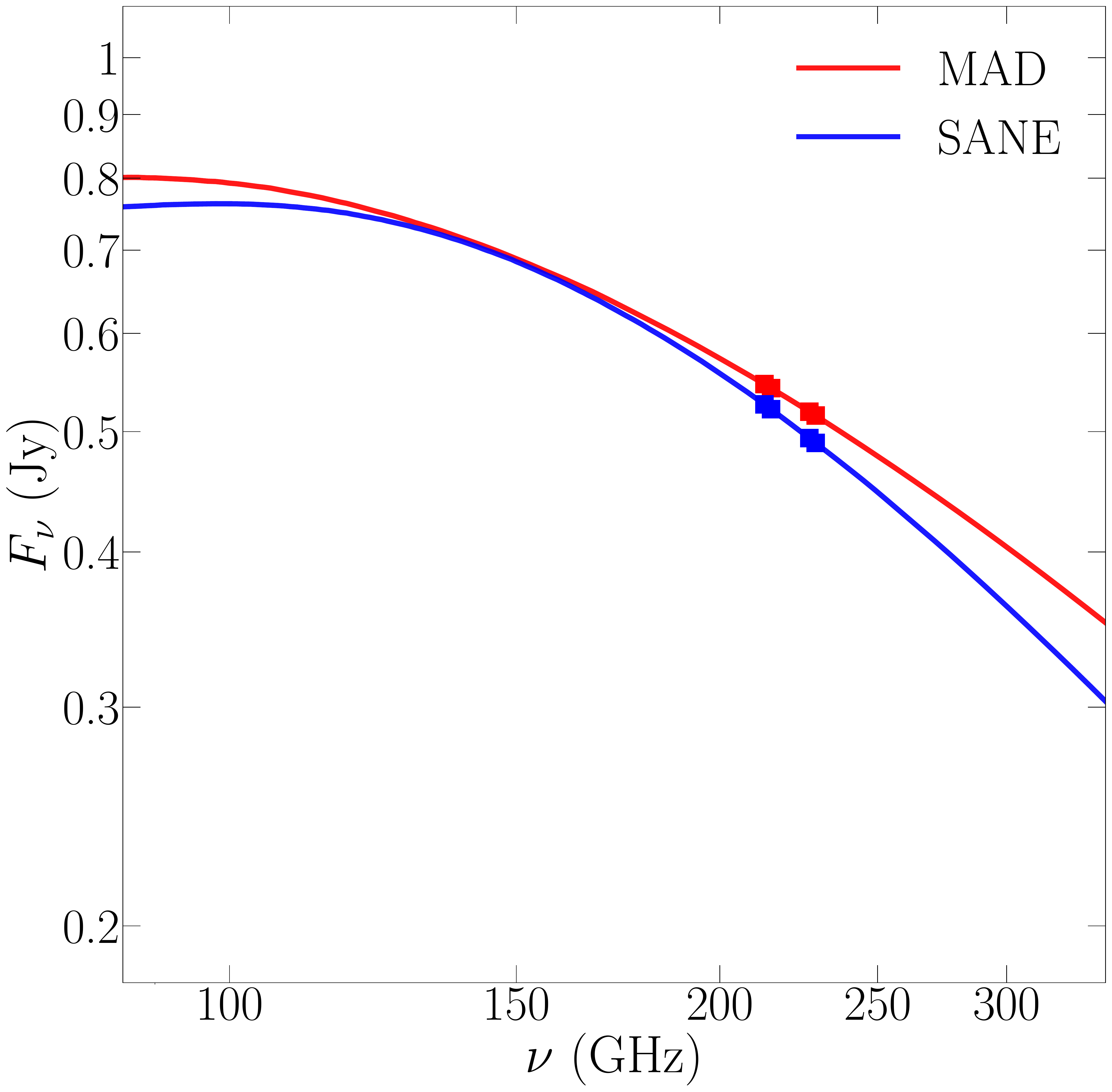}

\caption{(Left) $u-v$ coverage of the ngEHT concept array considered here for the four 230 GHz sub-bands at 213,215,227, and 229 GHz. (Right) synchrotron spectra of the two GRMHD simulation models considered here from \citet{Ricarte_2022}, indicating the points sampled by the four EHT bands.}
\label{fig:angeloUV}
\end{figure*}

\begin{figure*}[t]
\centering
\includegraphics[width=\textwidth]{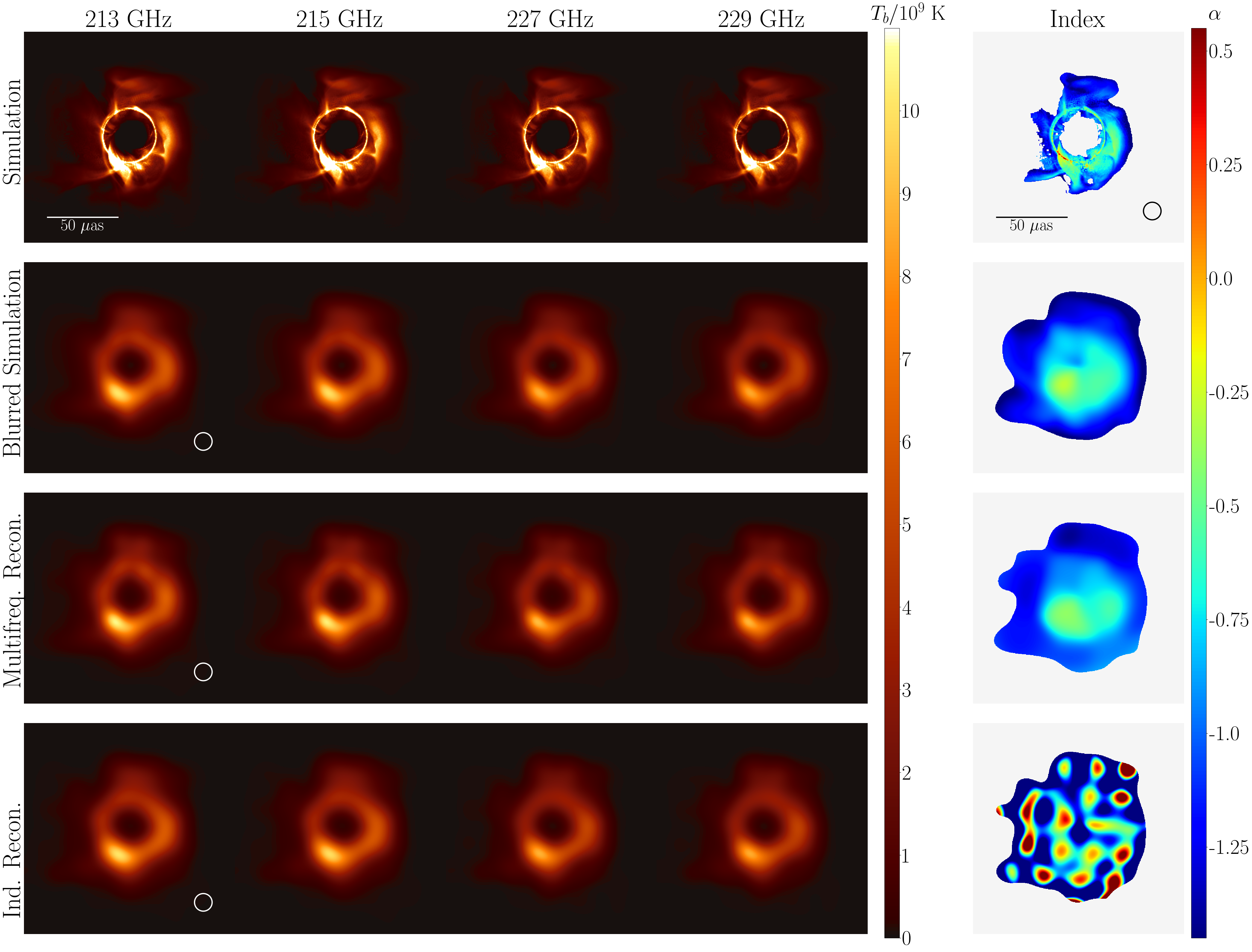}
\caption{Reconstructions of images of a fiducial MAD model of M87* from \citet{Ricarte_2022}. Columns from left to right show images and reconstructions at 213,215,227,and 229 GHz. The rightmost column shows the spectral index fit between the corresponding images in each row. The top row shows the simulation images at `infinite resolution'. The second column shows the ground truth simulation images blurred to 1/2 of the ngEHT nominal resolution at 1.3 mm. The third row shows reconstructions of these images from ngEHT data using multi-frequency synthesis, combining information across all four bands. The last row shows images reconstructed independently at each frequency band from the same data as the images in the third row. Although the independent reconstructions look to be similarly accurate to the multi-frequency reconstruction by eye, small offsets in the images in position and intensity mean they cannot be used to derive a 230 GHz spectral index map at this scale. By contrast, the multi-frequency synthesis approach in this paper accurately recovers the ground truth spectral index from 213-229 GHz. }
\label{fig:angeloMAD}
\end{figure*}

\begin{figure*}[t]
\centering
\includegraphics[width=\textwidth]{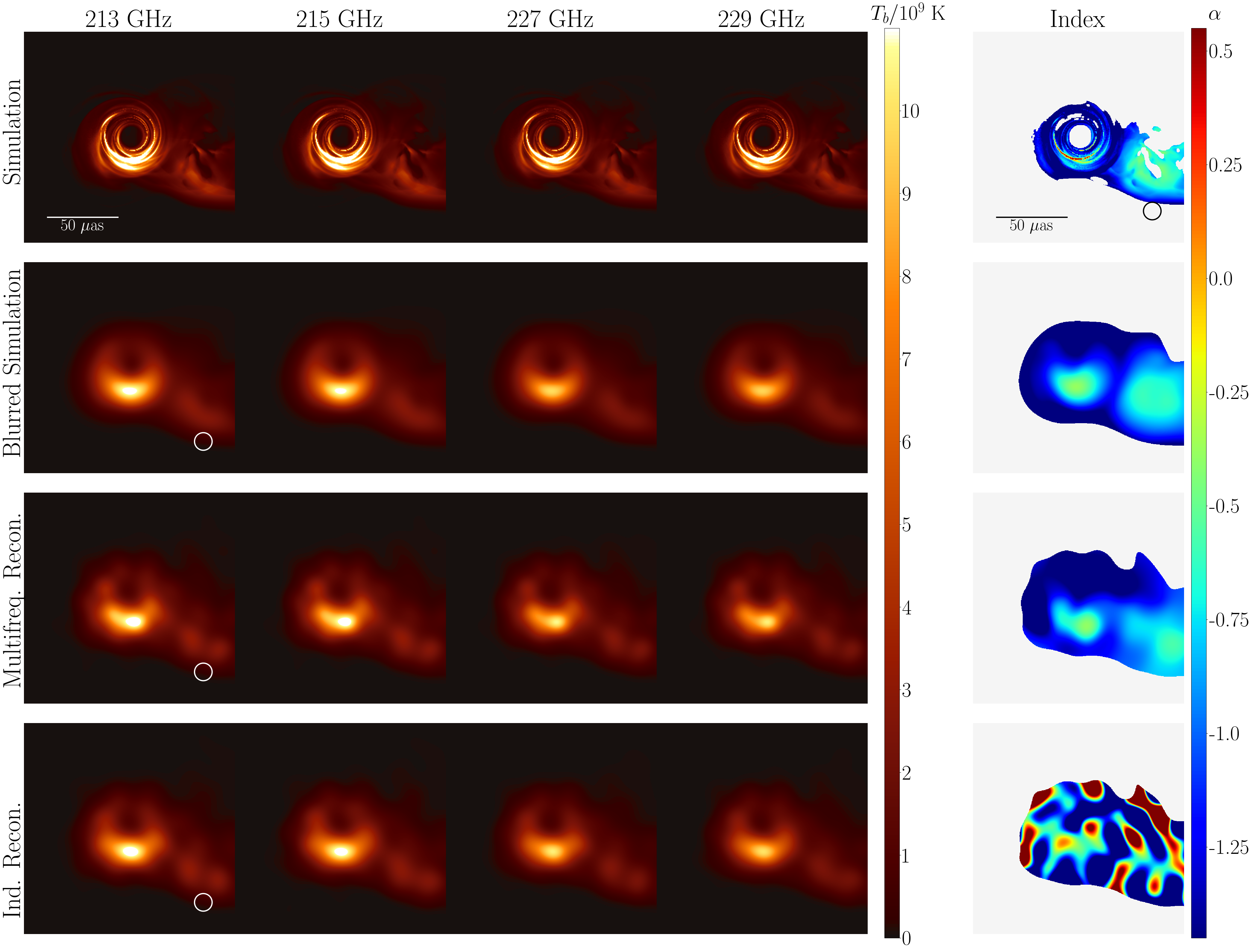}
\caption{Same as \autoref{fig:angeloMAD}, but for a fiducial SANE model of M87*. Note the differences in the underlying emission location and spectral index values from \autoref{fig:angeloMAD}. Spectral index maps can be used as a potential signature of different accretion states in resolved low-luminosity AGN images \citep{Ricarte_2022}.}
\label{fig:angeloSANE}
\end{figure*}

In this section, we consider reconstructions from simulated ngEHT data taken from two GRMHD models of M87* studied in \citet{Ricarte_2022}. In this example, we neglect 86 and 345 GHz observations and focus on the capabilities of the ngEHT to reconstruct resolved spectral structure over the narrower frequency band around 1.3 mm. The four ngEHT sub-bands are centered at 213, 215, 227, and 229 GHz; each has a bandwidth of 2 GHz. In the left panel of \autoref{fig:angeloUV}, we show the ngEHT \uv coverage for M87* across these four bands.

Our two source models from \citet{Ricarte_2022} are a MAD simulation with a black hole spin $a_* = -0.5$ and a SANE simulation with $a_*=0.94$. Both models were generated with the GRMHD code {\sc kharma} \citep[Prather et al. in prep.,][]{Prather_2021}, and synchrotron radiative transfer was performed with {\sc ipole} \citep{Moscibrodzka&Gammie2018}.  Both models have been scaled to produce approximately 0.5 Jy in flux at 230 GHz using a purely thermal electron distribution function. We take the $R_\mathrm{high}$ parameter that sets the ion-to-electron temperature ratio in the disk  \citep{Moscibrodzka_16M87} to be 160 in the MAD model and 10 in the SANE model.  As discussed in \citet{Ricarte_2022}, GRMHD models generically have a radially decreasing spectral index at these frequencies, due to declining optical depth, magnetic field strength, and temperature.  Equivalently, images grow smaller as frequency increases.  Constraining the amplitude of this radial decline may therefore be useful for constraining the radial evolution of these plasma parameters, which can vary by orders of magnitude among different models.

The right panel of \autoref{fig:angeloUV} shows the sub-millimeter spectra of these models, and indicates the four 1.3mm sub-bands considered in this study.  In both of these models, the overall spectral index is negative, as is the curvature, as we expect for synchrotron sources in the optically thin limit.  However, the spectral curvature is not significant over the range of frequencies covered by the four 1.3mm ngEHT sub-bands. As a result, we fix $\boldsymbol\beta=0$ in the reconstructions considered here and only reconstruct the spectral index map $\boldsymbol\alpha$.

For both models, we generate synthetic data on ngEHT baselines over the four sub-bands using the procedure described in \autoref{sec:syndata}. These data contain thermal noise and systematic amplitude and phase errors. We reconstruct images at each frequency independently using a standard RML method, and we also fit the data simultaneously in a multifrequency reconstruction for a reference frequency image $\mathbf{I}_0$ and spectral index map $\boldsymbol\alpha$. In both cases, we first fit the closure amplitude and closure phase data directly to account for the systematic gain and phase errors. We then self-calibrate the data to the output image at each frequency and re-image using the calibrated visibility amplitudes and the closure phases. Aside from the hyperparameters on the total variation and $\ell_2$ regularizers on the spectral index map $\boldsymbol\alpha$ in \autoref{eq:objfuncI}, we use the same hyperparameters for the data and regularizer terms for both the independent frequency reconstructions and the multi-frequency fit.

We show results of this test in \autoref{fig:angeloMAD} (for the MAD model) and \autoref{fig:angeloSANE} (for the SANE model). In each figure, the left four columns show the ground truth images or reconstructions from the simulated ngEHT data for the four sub-bands. The rightmost column shows the ground truth spectral index map, the spectral index map fit in the multi-frequency reconstruction (second to last row), and the spectral index map computed from  the independent reconstructions (last row). For the multifrequency RML reconstruction, the spectral index map is obtained directly as an output from the imaging process. For the ground truth and independent single-frequency reconstructions, the spectral index is obtained in each pixel through a linear fit.

In both examples, for both independent imaging at each frequency and in the multi-frequency fit, the recovered total intensity images reproduce the ground truth image structure well when blurred with a 12 $\mu$as kernel (1/2 the nominal resolution of the ngEHT at 1.3mm). However, because the full frequency range considered in this example is only 18 GHz, even small errors in the recovered intensity in the independent reconstructions translate to large errors in the recovered spectral index map. As a result, the spectral index maps recovered from the independent images do not accurately reproduce any physically useful information from the ground truth spectral index distribution at 1.3 mm.
In contrast, because of the combined constraints of the spectral index model \autoref{eq:loglog2} and the regularizing terms on the value and smoothness of  $\boldsymbol\alpha$ in \autoref{eq:objfuncI}, the simultaneous multi-frequency fit of all four datasets is able to accurately recover a spectral index map similar to the ground truth distribution, even though the individual images are nearly indistinguishable from the single-frequency reconstructions by eye.

Note that the poor spectral index map recovery in the single-frequency reconstructions in \autoref{fig:angeloMAD} and \autoref{fig:angeloSANE} is not a consequence of image-misalignment in the single-frequency results; we tested several strategies for image alignment, and we see similar results when we reconstruct images using complex visibilities in the synthetic data without any systematic phase or gain errors. Instead, the poor performance of the spectral index recovery from these single-frequency reconstructions is due to small errors in the independently recovered intensity maps at each frequency  translating to large errors in the spectral index over the small bandwidth. In this example, for context, differences in the image pixel intensities of only 5 percent between the smallest and largest frequency in this example correspond to a spectral slope of 0.7.

\begin{figure*}[t]
\centering
\includegraphics[width=\textwidth]{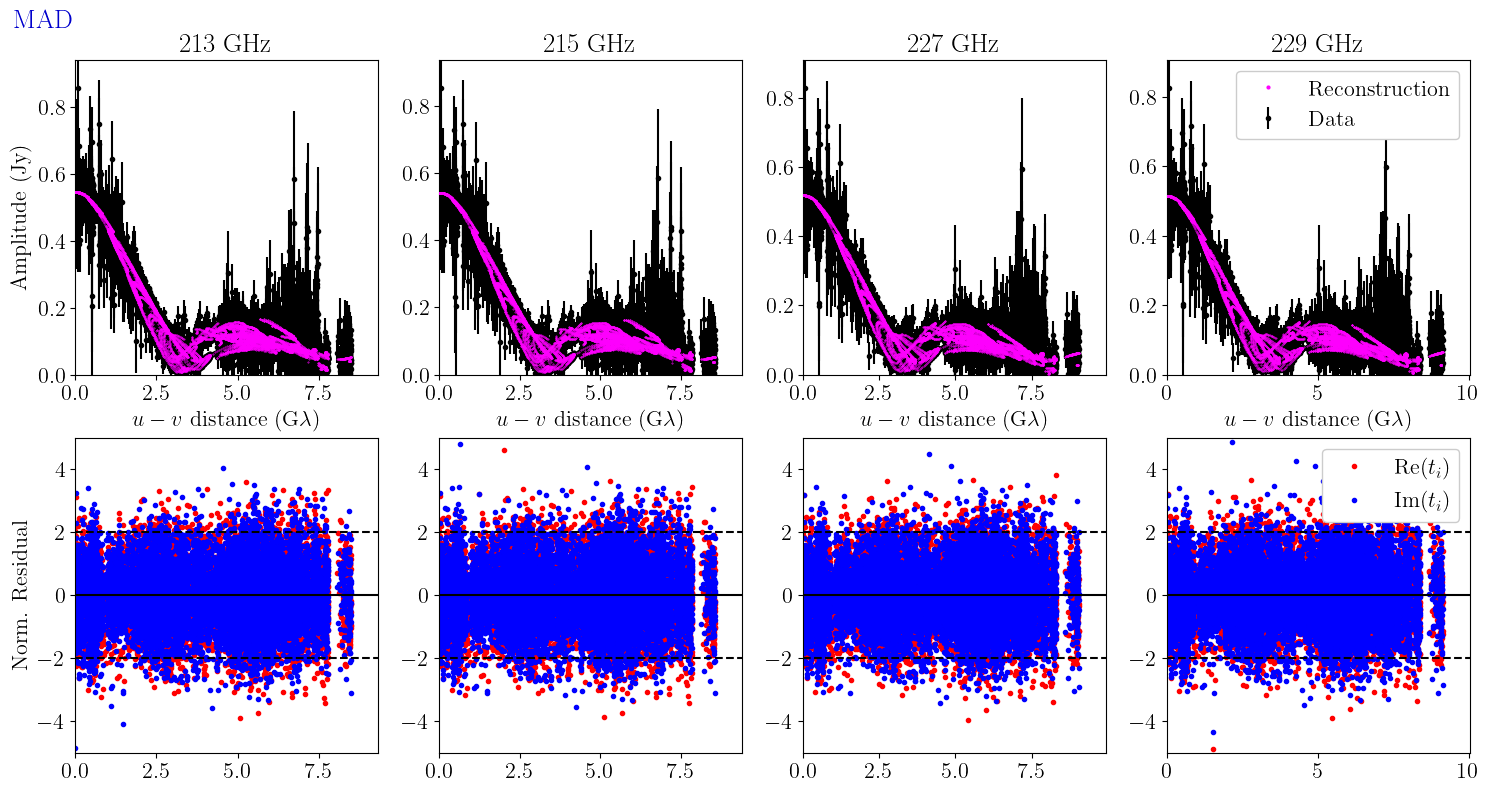}\\
\includegraphics[width=\textwidth]{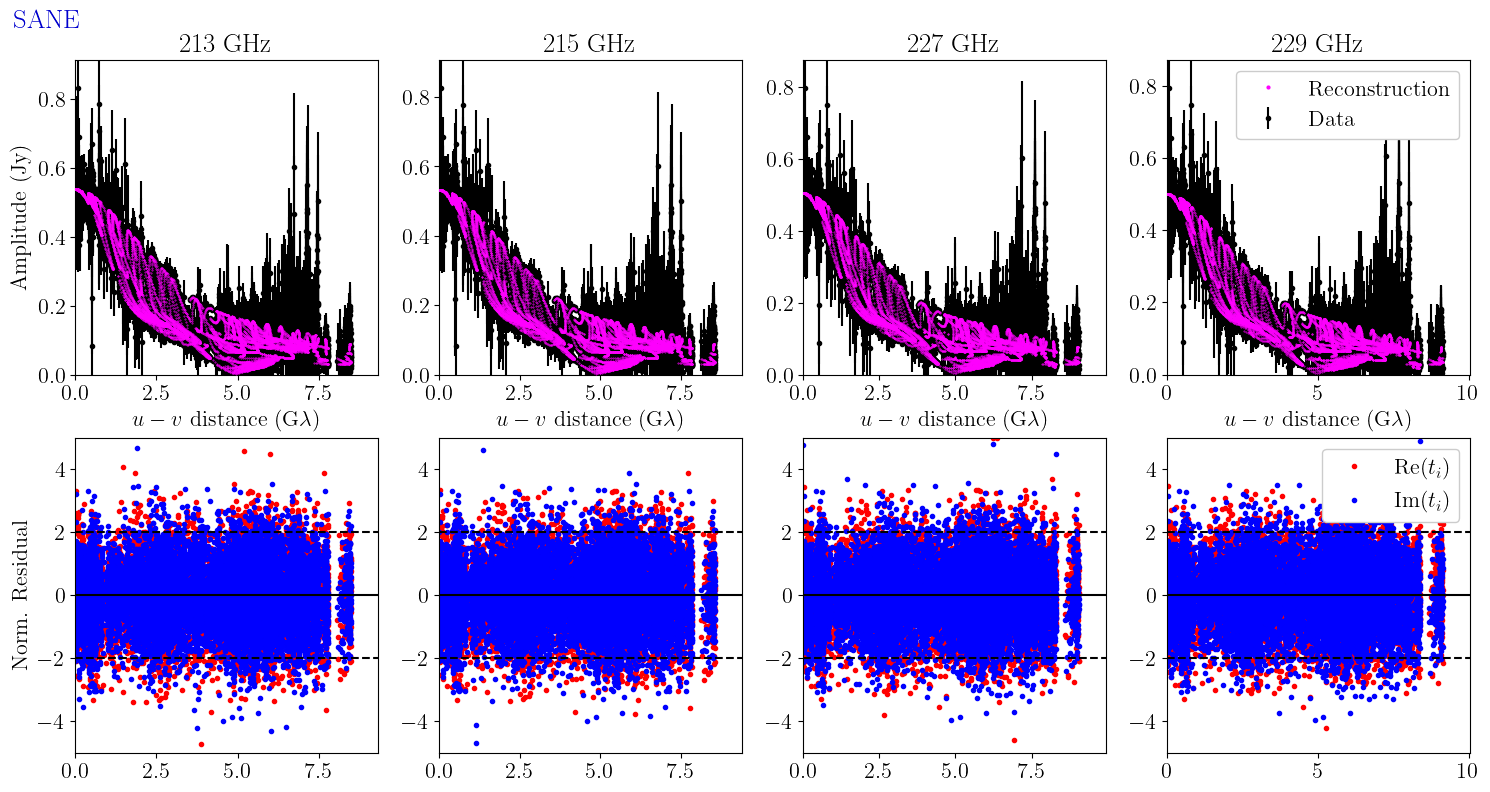}
\caption{(Top row) Comparison of the visibility amplitudes of the final self-calibrated data from the MAD model (black) to the model predictions from the final multi-frequency images (magenta) as a function of \uv distance. (Second row) Normalized complex residuals $t = (V_{\rm sc} - V_{\rm image})/\sigma^2$, where $V_{\rm sc}$ are the self-calibrated visibilities and $V_{\rm image}$ are the model visibilities predicted by the final multi-frequency image fit. We plot $\rm{Re}(t)$ in red and $\rm{Im}(t)$ in blue as a function of \uv distance.  (Third row) Comparison of the self-calibrated visibility amplitudes to the multi-frequency image reconstructions from the SANE model. (Fourth row) Normalized residuals from the multi-frequency image reconstructions of the SANE model.}
\label{fig:angeloResiduals}
\end{figure*}

In \autoref{tab:chisqs_snyth}, we present the reduced-$\chi^2$ statistics quantifying the fit quality for both the MAD (\autoref{fig:angeloMAD}) and SANE (\autoref{fig:angeloSANE}) models. As in \autoref{sec:ngEHT}, we find that both single- and multi-frequency image reconstructions give $\chi^2$ values close to unity, and we cannot distinguish between the different reconstruction methods using these goodness-of-fit statistics alone. In \autoref{fig:angeloResiduals} we illustrate the fit to data of the multifrequency image results for both the MAD and SANE models by comparing the visibility amplitudes of the data and reconstructed images across all four bands. We also show the normalized complex residuals of the reconstructed images compared to the self-calibrated visibility data; these residuals appear structureless and consistent with a unit normal distribution.

\section{Example Reconstructions: Real VLBA and ALMA data}
\label{sec:realdata}

\begin{figure*}[t]
\centering
\includegraphics[width=\textwidth]{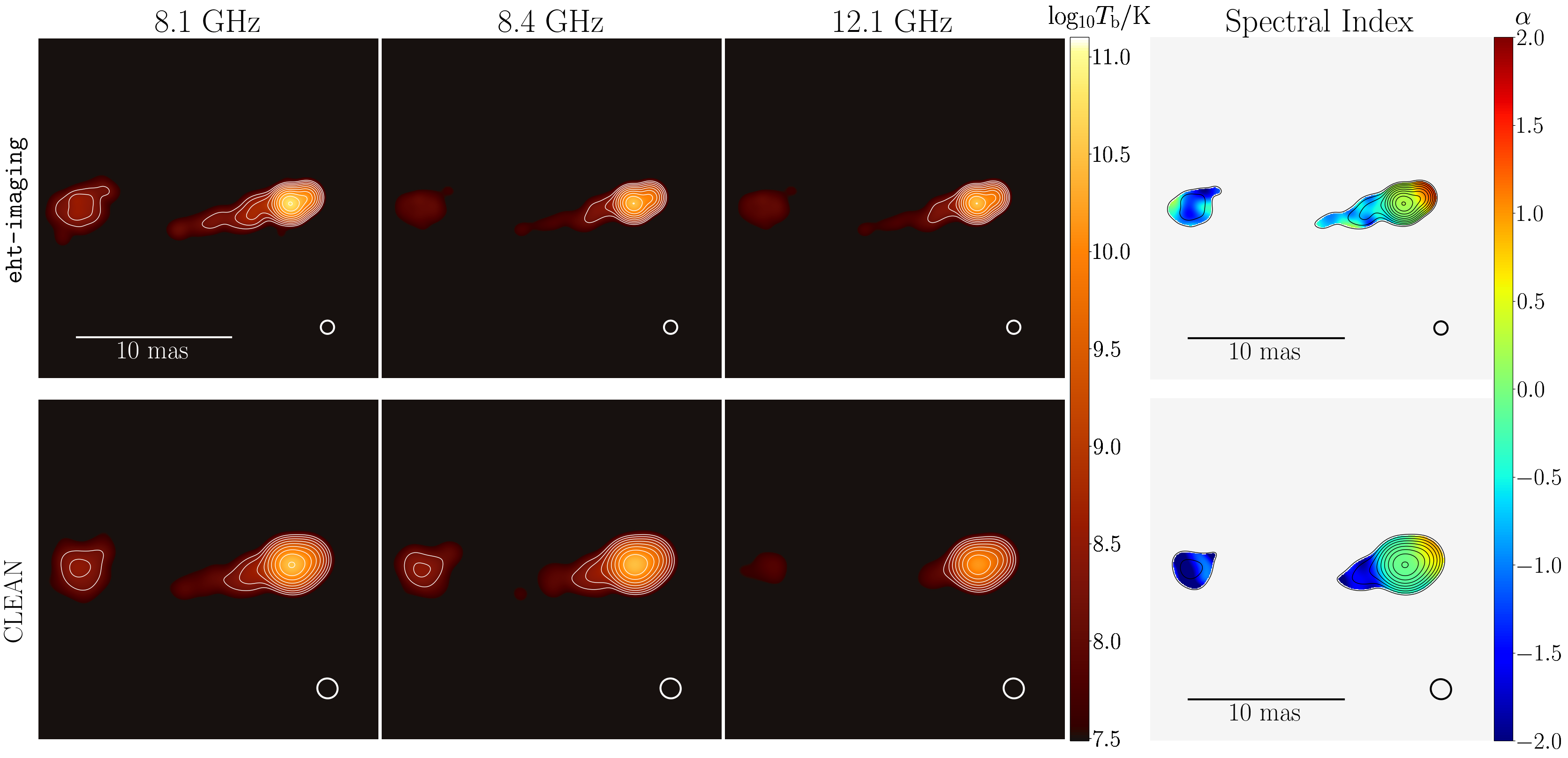}
\caption{Multi-frequency reconstructions of simultaneous VLBA observations of S5 0212+73 taken on July 7 2006 as part of the MOJAVE program \citep{Hovatta_2014,Lister2018}. The left three columns show image reconstructions at 8.1, 8.4 and 12.1 GHz, respectively; the rightmost column shows the spectral index. The top row shows simultaneous \texttt{eht-imaging} reconstructions of all three datasets performed with the method presented in this paper. The bottom row shows the original CLEAN reconstructions presented in \citep{Hovatta_2014}. The \texttt{eht-imaging} reconstructions are convolved with a circular Gaussian beam corresponding to the nominal resolution at 8.4 GHz; the original CLEAN reconstructions are convolved with the fitted CLEAN beam at 8.1 GHz used in \citet{Hovatta_2014}. Contours are drawn in steps of $1/2$ down from the peak brightness.
}
\label{fig:0212}
\end{figure*}

\begin{figure*}[t]
\centering
\includegraphics[width=\textwidth]{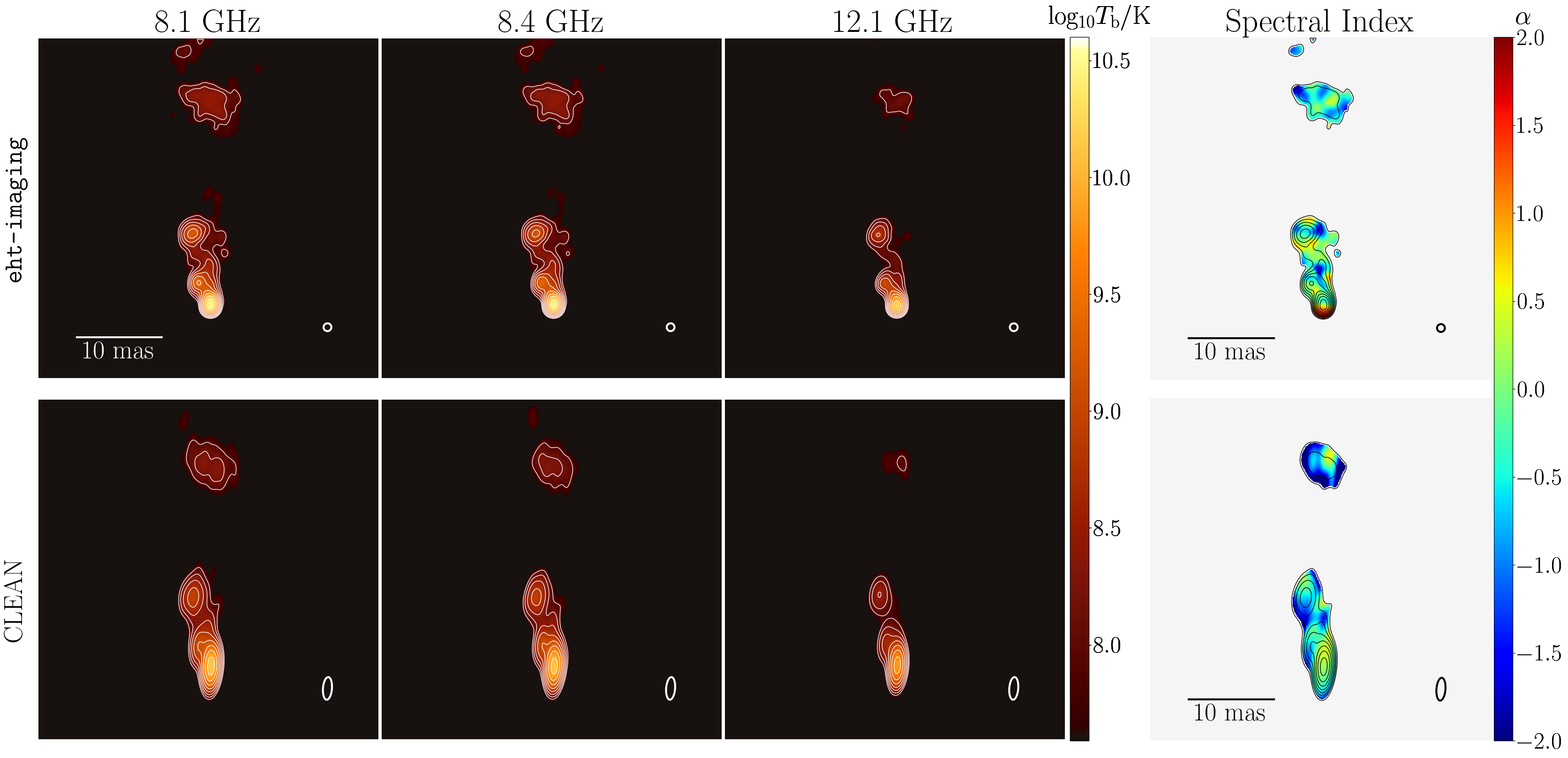}
\caption{Same as \autoref{fig:0212}, but for multi-frequency reconstructions of simultaneous MOJAVE VLBA observations of NRAO~530 (1730-130) taken on July 7 2006 \citep{Hovatta_2014,Lister2018}.}
\label{fig:1730}
\end{figure*}

\begin{figure*}[t]
\centering
\includegraphics[width=\textwidth]{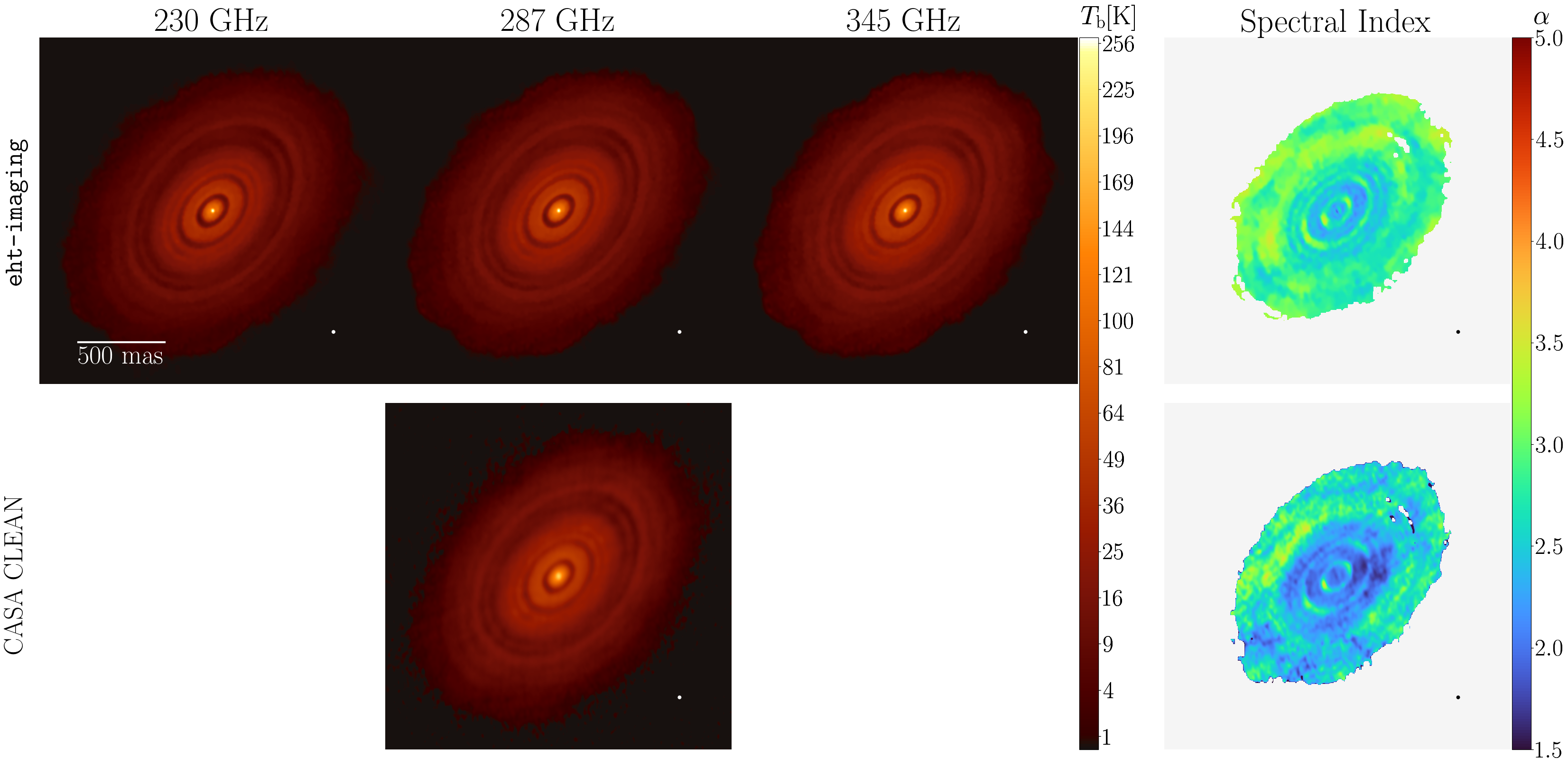}
\caption{Multi-frequency reconstructions of ALMA observations of HL Tau across 8 spectral windows in Band 6 and Band 7. The first and third panels show the reconstruction at 230 (Band 6) and 345 GHz (Band 7), respectively. The second panel shows reconstructions at 287 GHz, the central frequency used in the original CASA CLEAN multi-frequency reconstruction in \citep{ALMA_HlTau}. The fourth column shows the spectral index map. The top row shows the results of \texttt{eht-imaging} reconstructions performed on the eight spectral window datasets assuming no spectral curvature. The bottom row shows the original CASA CLEAN reconstructions and spectral index map. Because the released image files from \citep{ALMA_HlTau} does not include spectral index values outside the masked region shown, it is not possible to fully reconstruct the 230 and 345 GHz images from the available 287 GHz image and spectral index map.}
\label{fig:HLTau}
\end{figure*}

In this section, we present three examples of applying the simultaneous RML spectral index imaging method described in this paper to real interferometric data sets. In \autoref{sec:mojave} we consider two examples of VLBI spectral index imaging using MOJAVE data from $8-12$ GHz \citep{Hovatta_2014,Lister2018}. In \autoref{sec:hltau} we demonstrate the applicability of the method to larger datasets from connected-element interferometers with a reconstruction of the spectral index structure of the protoplanetary disk in HL Tau from 2014 ALMA observations between $224-350$ GHz \citep{ALMA_HlTau}.\footnote{The reduced data and \texttt{eht-imaging} scripts used to produce the reconstructed images in the following sections can be found at https://github.com/achael/multifrequency\_scripts/.}

\subsection{VLBA imaging of MOJAVE data sets}
\label{sec:mojave}

In \autoref{fig:0212} and \autoref{fig:1730} we present multi-frequency reconstructions of two jet sources observed at 8.1, 8.4, and 12.1 GHz as part of the MOJAVE program\footnote{https://www.cv.nrao.edu/MOJAVE/index.html} in July 2006 \citep{Hovatta_2014,Lister2018}. In \autoref{fig:0212} we show image and spectral index reconstructions for the BL Lac source S5 0212+73;\footnote{https://www.cv.nrao.edu/MOJAVE/sourcepages/0212+735.shtml}
in \autoref{fig:1730} we show image and spectral index reconstructions for the blazar NRAO530.\footnote{https://www.cv.nrao.edu/MOJAVE/sourcepages/1730-130.shtml}

In both cases, we produced initial images from these datasets by including only closure amplitudes and closure phases in the log-likelihood terms in the objective function $J$ (\autoref{eq:objfuncI}). We then enforced the total flux density at each frequency through self-calibration of the visibility amplitudes using the total flux densities taken from the publicly available MOJAVE images reconstructed with CLEAN. We then re-imaged using visibility amplitude and closure phase terms in the log-likelihood part of the objective function. Throughout, we included $\ell_1$ (sparsity) and total variation (smoothness) regularizers on the reference frequency image (at 8.4 GHz) and we included a total variation regularizer on the spectral index map. In both sources, the spectral curvature over the observed frequency range is minimal, so we reconstructed only the first-order spectral index map $\boldsymbol\alpha$ and set the spectral curvature term $\boldsymbol\beta=0$.

In \autoref{fig:0212} and \autoref{fig:1730} we show results for both our simultaneous \texttt{eht-imaging} RML reconstructions and the original CLEAN reconstructions presented in \citet{Hovatta_2014}, which were preformed independently at the three frequencies and then aligned. We convolved the \texttt{eht-imaging} reconstructions with a circular Gaussian beam corresponding to the nominal resolution at 8.4 GHz; in contrast, the original CLEAN results are presented after convolution with the anisotropic CLEAN beam at 8.1 GHz, as presented in \citet{Hovatta_2014}. In \autoref{tab:chisqs_mojave}, we present reduced-$\chi^2$ statistics of our multi-frequency image reconstructions for the log closure amplitudes, closure phases, and visibility amplitudes.

In both sources, our results reproduce essential features of the original spectral index maps from \citet{Hovatta_2014}. In both \autoref{fig:0212} and \autoref{fig:1730}, there is a clear decline in the spectral index from positive values to negative values along the jet with increasing distance from the core. The spectral index map  recovered in our reconstructions of the NRAO 530 observations (\autoref{fig:1730}) shows more structure on the scales of the observing resolution than are seen in the reconstructions of S5 0212+735 (\autoref{fig:0212}), including patches of positive spectral index downstream of the core; similar features are also seen in the CLEAN reconstructions, though on larger scales corresponding to their larger restoring beam.

The recovered spectral index far from the core in S5 0212+735 (\autoref{fig:0212}) is more negative in the original CLEAN reconstructions of these datasets (where it reaches values of $\alpha\sim-2.5$) than in the \texttt{eht-imaging} reconstruction (where the lowest values are $\alpha\sim-1.5$). To see if this preference for larger spectral index values in the extended jet was an artifact of our imaging choices, we experimented with several choices of regularizing terms in the objective function (\autoref{eq:objfuncI}), including an $\ell_2$ norm term (\autoref{eq:sell2}) that preferred a strongly negative spectral index $\alpha=-2.5$ in the absence of data constraints. None of these reconstructed images with different regularizer choices gave substantially different values for the downstream spectral index in the \texttt{eht-imaging} reconstructions.

\subsection{ALMA imaging of HL Tau}
\label{sec:hltau}

In \autoref{fig:HLTau}, we present multi-frequency reconstructions of observations of the protoplanetary disk in HL Tau conducted as part of the ALMA science verification process\footnote{https://almascience.nrao.edu/alma-data/science-verification/science-verification-data} and published in \citet{ALMA_HlTau}. We reconstructed multifrequency images from the publicly available ALMA datasets across the four spectral windows in both ALMA Band 6 (centered on 241 GHz) and Band 7 (centered at 324 GHz). Before imaging, we reduced the data by averaging the complex visibilities in frequency across the eight spectral windows and in time windows of 300 seconds to reduce the data volume.

In \texttt{eht-imaging}, we fit the eight spectral window datasets from  Band 6 and Band 7 simultaneously to an image model and spectral index map $\boldsymbol\alpha$ (setting the spectral curvature $\boldsymbol\beta=0$).
Because the data volume remains much larger than the VLBI datasets we typically reconstruct with \texttt{eht-imaging}, for this image reconstruction we used use complex visibilities directly in the likelihood terms of the objective function (\autoref{eq:objfuncI}) instead of closure quantities. We start with the initial amplitude and phase calibration provided in the public ALMA data, but we self-calibrate all eight spectral window datasets three times to intermediate results during the imaging process. For regularizing terms in \autoref{eq:objfuncI}, we use maximum entropy and total variation on the reference image (at 287 GHz) as well as total variation  on the spectral index map (\autoref{eq:alphatv}) and an $\ell_2$ norm term on the spectral index map (\autoref{eq:sell2}) with a fiducial spectral index value $\alpha_0=0$.

In \autoref{fig:HLTau} we present the resulting images (at 230, 287, and 345 GHz) and the spectral index map from our \texttt{eht-imaging} reconstruction. We compare our results to the original images presented in \citet{ALMA_HlTau}, which were conducted with multi-frequency CLEAN imaging in CASA \citep{Rau_2011}.\footnote{https://almascience.nrao.edu/almadata/sciver/HLTauBand7/}  In \autoref{tab:chisqs_hltau}, we present reduced-$\chi^2$ statistics of our multi-frequency image reconstructions for the complex visibilities from all eight bands, which we fit directly in this example.

Our results are broadly consistent with the published images from \citet{ALMA_HlTau}. We obtain a clear concentric ring structure with distinct gaps between adjacent rings. In the spectral index maps, these gaps correspond to local maxima in the spectral index, whereas the bright ring structures have smaller values of the spectral index. Compared to the original CLEAN reconstruction, the \texttt{eht-imaging} reconstruction produces sharper spectral features corresponding to the intermediate dark rings \citep[rings D3 and D4 in][]{ALMA_HlTau}.

The values of the spectral index we recover throughout the image are somewhat higher than in the original CLEAN reconstructions from \citet{ALMA_HlTau}.
This discrepancy is likely caused by differences in the overall amplitude calibration in our reconstructions as compared to the original CLEAN images.
During the \texttt{eht-imaging} procedure, we rescale the total flux density of the intermediate images used for self-calibration to the total flux density values reported in Table 1 of \citet{ALMA_HlTau}. This procedure produces an unresolved spectral index $\alpha=2.74$ over the band from 224 to 351 GHz, which matches the reported unresolved spectral index of $\alpha=2.77\pm0.13$ in \citet{ALMA_HlTau}.
By contrast, the publicly available multi-frequency CLEAN images from the Band 6+7 reconstruction in \citet{ALMA_HlTau} have an unresolved spectral index of only $\alpha=2.36$. If we instead set the total flux density for amplitude calibration in our \texttt{eht-imaging} procedure to values taken from the multi-frequency CLEAN image, we obtain lower spectral index values throughout the image that more closely match the publicly available CLEAN images.

\begin{table*}[ht]
    \centering
    \begin{tabular}{c|cc|ccc}
    \hline
      Source  & Method & Frequency  & $\chi^2_{\mathrm{log} A_C}$ & $\chi^2_{\psi_C}$ & $\chi^2_{|V|}$  \\
    \hline \hline
     S5 0212+73 (\autoref{fig:0212}) & Multi-Frequency & 8.1 GHz & 1.05 & 1.17 & 0.86 \\
       & & 8.4 GHz & 1.09 & 1.09  & 0.88  \\
       & & 12.1 GHz & 1.32  & 1.32 & 1.06 \\
    \hline
     NRAO 530 (\autoref{fig:1730}) & Multi-Frequency & 8.1 GHz & 1.08 & 1.20 & 0.83 \\
       & & 8.4 GHz & 0.60 & 1.41 & 0.34 \\
       & & 12.1 GHz & 1.14 & 1.05 & 0.85  \\
       \hline
    \end{tabular}
    \caption{Reduced $\chi^2$ table for MOJAVE multi-frequency RML image reconstructions. For the multi-frequency reconstructions of S5 0212+73 (\autoref{fig:0212}) and NRAO530 (\autoref{fig:1730}), we provide reduced-$\chi^2$ statistics comparing the final self-calibrated data sets for each frequency band with the final multi-frequency RML image reconstructions for the data products used in the fit: log closure amplitudes ($\chi^2_{\mathrm{log}A_{\mathcal{C}}}$), closure phases ($\chi^2_{\psi_{\mathcal{C}}}$), and visibility amplitudes ($\chi^2_{|V|}$).}
    \label{tab:chisqs_mojave}
\end{table*}

\begin{table*}[ht]
    \centering
    \begin{tabular}{c|cc|c}
    \hline
      Source  & Method & Frequency  & $\chi^2_{V}$  \\
    \hline \hline
     HL Tau (\autoref{fig:HLTau}) & Multi-Frequency & 224 GHz & 1.07\\
     & & 226 GHz & 1.05 \\
     & & 240 GHz & 1.11 \\
     & & 242 GHz & 1.07 \\
     & & 336 GHz & 0.95 \\
     & & 338 GHz & 0.95 \\
     & & 345 GHz & 1.01 \\
     & & 351 GHz & 1.05 \\
     \hline
    \end{tabular}
    \caption{Reduced $\chi^2$ table for the ALMA multi-frequency RML image reconstruction in \autoref{fig:HLTau}. We provide reduced-$\chi^2$ statistics comparing the final self-calibrated data sets for each frequency band with the final multi-frequency RML image reconstructions for the complex visibilities only, since in this example we fit directly to complex visibilities instead of closure quantities and visibility amplitudes.}
    \label{tab:chisqs_hltau}
\end{table*}

\section{Discussion}
\label{sec:Discussion}

Our method for multi-frequency RML imaging has several distinct advantages over more commonly used CLEAN-based algorithms for multi-frequency synthesis \citep{Rau_2011,Offringa_2021}. Most importantly, as an RML-based method, our approach can directly reconstruct images using robust interferometric `closure' products; the likelihood terms in the objective function (\autoref{eq:objfuncI}) can be constructed with any data product derived from the observations. In contrast, all CLEAN-based image reconstruction algorithms require an initial calibration step so that the data can be inverse Fourier-transformed to produce a dirty image, and a poor specification of the initial self-calibration model can lead to large image errors. In our approach, even when self-calibration is used, the initial self-calibration model can always be derived from an initial fit to the most robust data products (in our case, closure phases and log closure-amplitudes). In this paper, only in the reconstruction of ALMA HL Tau observations (\autoref{fig:HLTau}) did we fit directly to complex visibilities at any stage of the imaging process. The RML method's ability to fit directly to closure phases and closure amplitudes makes it  naturally suited for imaging millimeter and submillimeter VLBI data sets. Gain calibration is difficult at these high frequencies, and atmospheric fluctuations typically make recovering absolute visibility phase nearly impossible.

When deriving a spectral index map from interferometric images reconstructed without absolute phase information, determining the relative alignment of the images is a major source of systematic error \citep[e.g.][]{Hovatta_2014}. We found that RML multifrequency synthesis performed well in all of the example data sets reconstructed in this paper without absolute phase information; the RML imaging process did not introduce any clear artifacts from image misalignment in any of these examples. RML multifrequency synthesis effectively enforces image alignment by the spectral index regularization terms, which favor spectral index maps without large gradients (\autoref{eq:alphatv}) and where the spectral index remains close to the unresolved value (\autoref{eq:sell2}). Simultaneous RML reconstruction does not completely remove the possibility for image misalignment between frequencies; position offsets may be a particular problem for poorly-resolved images where there are few common features between frequency bands for the imager to `lock on' to during image reconstruction.

In addition to enforcing alignment between reconstructed images at different frequencies, RML  multifrequency reconstructions can obtain higher levels of image `superresolution' of features finer than the nominal resolution scale than is possible  the corresponding single-frequency reconstructions. For example, in the ngEHT jet reconstructions in figure \ref{fig:rjetI}, superresolution is enhanced in the multifrequency reconstruction by propagating structural information from the 230 GHz and 345 GHz data to the 86 GHz image, subject to regularizing constraints on the values and smoothness of the spectral maps. As a result, the multifrequency 86 GHz reconstruction can superresolve the central brightness depression around the central black hole at 86 GHz that are not seen in the single-frequency reconstruction.

Another advantage to the RML approach for multi-frequency synthesis is its flexibility. RML imaging can be easily adapted to a specific problem by modifying the objective function (\autoref{eq:objfuncI}). The \texttt{eht-imaging} code is adapted to this flexibility in the method; new likelihood terms or new regularizing terms can be developed and added to the objective function with minor alterations to the imaging code. In this paper we only use two regularizing functions on the spectral index and curvature maps that promote similarity to a fiducial value   (\autoref{eq:sell2}) and spatial smoothness (\autoref{eq:alphatv}), but new regularizers may easily be developed for future applications. For instance, it may be useful to apply prior information on the spatial variation of the spectral index map with a maximum-entropy term, or to regularize the spatial power spectrum of the reference frequency image or spectral index map.

The multi-frequency RML imaging approach can also be straightforwardly adapted for polarization imaging \citep[e.g.][]{Chael_16}. In a forthcoming work we will present and test a method for direct rotation-measure synthesis using an extension of the multifrequency RML techniques presented here. In this approach, we simultaneously fit images of the total intensity image $\mathbf{I}$, fractional polarization $\mathbf{m}$, polarization position angle $\boldsymbol\chi$ and rotation measure $\mathbf{R}$ to polarimetric multi-frequency data sets \citep[e.g.][]{Brentjens_2005,Andrecut_2012,Bell_2012}. Measuring and imaging the rotation measures of radio sources is particularly important in constraining the geometry of magnetic fields in relativistic jets  \citep[e.g][]{Gabuzda_2004,Hovatta_2012} and constraining the plasma density in accretion flows \citep[e.g.][]{Marrone_2007,Bower_2018}.

Despite its flexibility and good performance on simulated and real data, RML multi-frequency imaging also has several disadvantages to alternative image-reconstruction methods. The forward-modeling approach can become computationally inefficient when the number of observed data points is large.  For VLBI arrays like EHT and even the ngEHT, the \uv  coverage is sparse and the number of observations is small, so computing the image likelihood terms $\mathcal{L}(\mathbf{d}|\mathbf{I})$  (and their gradients) is fast. For observations from connected-element interferometers like ALMA, the large size of the data vector $\mathbf{d}$ slows down the likelihood and gradient computation; when imaging ALMA data in \texttt{eht-imaging}, we need to significantly average the data in time and frequency to enable image reconstruction in a reasonable time. In contrast, in CLEAN-based multi-frequency imaging methods \citep{Rau_2011,Offringa_2021}, the data is gridded once and inverse Fourier transformed before CLEANing. As a result, the size of the dataset only affects the speed of this initial step, and the speed of the iterative steps in the CLEAN loop are only affected by the image resolution, so CLEAN is well-adapted to very large interferometric data sets.

Another disadvantage of our current RML approach is that we fit images on a grid with a fixed pixel size; in contrast, existing CLEAN-based multi-frequency methods \citep{Rau_2011,Offringa_2021} currently employ multi-scale image bases \citep[e.g.][]{Cornwell_08}. As the source size can change with frequency, multi-scale approaches may be particularly important for simultaneous image reconstruction over a large range of frequencies. For example, in the ngEHT M87 reconstructions presented in \autoref{fig:rjetI}, the 86 GHz observation is more sensitive to large-scale structure in the extended jet, while the 345 GHz observation is most sensitive to fine-scale structure in the core close to the black hole. To fit both of these regimes, we require a large number of pixels across the image; it would be more efficient to be able to adapt the pixel resolution to the scale of the structures in these regions. Furthermore, we may wish to have different resolutions in the reference frequency image and the spectral index map if, for instance, we have a priori reason to expect the spectral index to vary more smoothly across the image than the total intensity image structure. Adding multi-scale image bases to the RML imaging code in \texttt{eht-imaging} will be a key next step to make the method more widely applicable to interferometric imaging in different regimes, and for more accurate and efficient imaging across wider frequency bands.

Finally, another key disadvantage of our RML multi-frequency synthesis technique is that we do not provide any robust measurement of the  uncertainty in the reference frequency image or in the spectral index map. Traditional CLEAN imaging methods use the Fourier transform of the image residuals as an estimate of the image uncertainty or noise \citep{Hog_1974,Clark_80}; CLEAN-based multi-frequency synthesis algorithms extend that approach and use the residual maps at different frequencies to estimate the uncertainty in the spectral index map \citep{Rau_2011}. However, the residual map is a poor estimate of image uncertainty (and dynamic range) when the \uv  coverage is sparse and when systematic uncertainty in the amplitude and phase calibration dominate over thermal noise, as is the case for the EHT \citep[e.g.][]{Cornwell_1981,Pearson_1984,Cornwell_1999,PaperIV}. A better approach in quantifying uncertainty in an interferometric image is to extend RML imaging to perform full Bayesian modeling of the image pixels. Instead of finding just a single maximum of the regularized likelihood function (\autoref{eq:objfuncI}), Bayesian inference techniques \citep[e.g.][]{Broderick_2020_Themis, Pesce_2021,Arras_2022}  can estimate full posterior probability distributions of the image pixels when fitting VLBI data sets. These Bayesian imaging algorithms have already been successfully used in fitting EHT observations of Sgr A* \citep{SgrAPaperIII} and M87* \citep{PaperVII}.

\section{Conclusion}
\label{sec:Conclusion}
In this paper we present a new method for multi-frequency image reconstruction of interferometric data sets. The method is a straightforward extension (\autoref{sec:method}) of existing regularized maximum likelihood imaging approaches now in common use for VLBI imaging with sparse arrays like the EHT \citep[e.g.][]{PaperIV,SgrAPaperIII}. We have implemented the method in the \texttt{eht-imaging} Python software library \citep{Chael_16,Chael18_Closure}. In our method, all of the existing data likelihood terms and total intensity image regularizers, imaging options, and calibration strategies in \texttt{eht-imaging} can also be used directly in multi-frequency synthesis; most importantly, we can still fit simultaneously to robust closure phases and closure amplitudes in generating multi-frequency reconstructions.

In \autoref{sec:ngEHT} we demonstrated that the method performs well at recovering spectral image structure from realistic simulated observations with a next-generation Event Horizon Telescope (\autoref{fig:ngEHT}) over a wide frequency range from 86 to 345 GHz. Simultaneous image reconstruction across frequencies is critical for robust recovery of spectral index and curvature information (\autoref{fig:rjetspec}, \autoref{fig:cjetspec}). In addition to naturally aligning images in reconstructions which may lack absolute phase information, simultaneous RML imaging can `share' information between frequencies and allow data at a given frequency to serve as an effective regularizer on the reconstruction at another frequency. We found this propagation of information across the band to be particularly important in recovering extended jet structure in simulated ngEHT images of M87* at 345 GHz, where the signal-to-noise is expected to be low (\autoref{fig:rjetI}, \autoref{fig:cjetI}). Simultaneous imaging from 86-345 GHz enhances the image resolution in 86 GHz images from simulated ngEHT data, allowing them to superresolve structure finer than the nominal 86 GHz resolution (\autoref{fig:superres}). This superresolution  is enough to directly image the central brightness depression at 86GHz. We also demonstrated that simultaneous multi-frequency RML imaging can recover accurate spectral index information even over the relatively small 18 GHz range between the lowest and highest ngEHT bands at 1.3 mm (\autoref{fig:angeloMAD}, \autoref{fig:angeloSANE}).

In \autoref{sec:realdata}, we demonstrated that our RML method can successfully reconstruct images and spectral index information from existing datasets from the VLBI and ALMA. While not identical to existing CLEAN reconstructions, our image reconstructions and spectral index maps of MOJAVE jet sources (\autoref{fig:0212}, \autoref{fig:1730}) and the HL Tau protoplanetary disk (\autoref{fig:HLTau}) reproduce the primary spatial and spectral features seen in prior CLEAN reconstructions. These results indicate that while RML multi-frequency synthesis will be critical for ngEHT imaging, it also has wide applicability in interferometry. In \autoref{sec:Discussion} we discussed advantages and disadvantages of RML imaging for multi-frequency image reconstruction. In future work we will extend our method to polarimetry and rotation-measure synthesis, and we will adapt it with multi-scale image bases to more efficiently reconstruct structure across a wide range of spatial scales, such as will be observed by the ngEHT in M87* and other sources.

\acknowledgments
The authors thank Paul Tiede for serving as the EHT
collaboration internal referee for this paper. The authors also thank the anonymous \emph{ApJ} referee for comments and suggestions that significantly improved the manuscript.
Matt Lister provided essential guidance and advice on the MOJAVE datasets used in \autoref{sec:mojave}.

AC was supported by Hubble Fellowship grant HST-HF2-51431.001-A awarded by the Space Telescope Science Institute, which is operated by the Association of Universities for Research in Astronomy, Inc., for NASA, under contract NAS5-26555.
SI is supported by Hubble Fellowship grant HST-HF2-51482.001-A awarded by the Space Telescope Science Institute, which is operated by the Association of Universities for Research in Astronomy, Inc., for NASA, under contract NAS5-26555. AC also acknowledges support from the Princeton Gravity Initiative.

Support for this work was provided by the NSF through grants AST-1440254, AST-1935980, and AST-2034306, and by the Gordon and Betty Moore Foundation through grant GBMF-10423.  This work has been supported in part by the Black Hole Initiative at Harvard University, which is funded by grants from the John Templeton Foundation and the Gordon and Betty Moore Foundation to Harvard University.

CMF is supported by the DFG research grant ``Jet physics on horizon scales and beyond" (Grant No.  FR 4069/2-1). YM is supported by the National Natural Science Foundation of China (Grant No. 12273022).
CMF and YM performed the simulations on LOEWE at the CSC-Frankfurt, Iboga at ITP Frankfurt and Pi 2.0 and Siyuan Mark-I at Shanghai Jiao Tong University.

This research has made use of data from the MOJAVE database that is maintained by the MOJAVE team \citep{Lister2018}.

This paper makes use of the following ALMA data:
ADS/JAO.ALMA\#2011.0.000015.SV.
ALMA is a partnership of ESO (representing its member states), NSF (USA) and NINS (Japan), together with NRC (Canada), MOST and ASIAA (Taiwan), and KASI (Republic of Korea), in cooperation with the Republic of Chile. The Joint ALMA Observatory is operated by ESO, AUI/NRAO and NAOJ.
The National Radio Astronomy Observatory is a facility of the National Science Foundation operated under cooperative agreement by Associated Universities, Inc.
\clearpage
\appendix
\section{Effect of Total Variation Regularization on Spectral Index Maps}

\begin{figure*}[h]
\centering
\includegraphics[width=\textwidth]{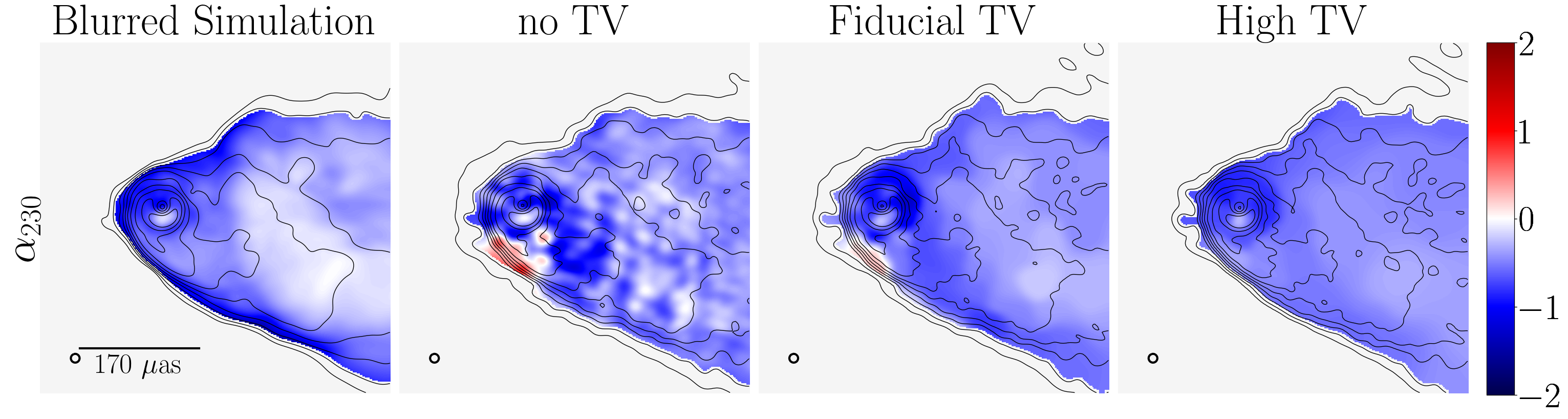}
\caption{Comparison of the recovered spectral index map from multi-frequency imaging of the simulation of \citet{ChaelM87} presented in \autoref{fig:rjetspec} with different values of the hyperparameters $\lambda_R$ weighting the total variation terms for spectral index ($S_{\rm TV}(\boldsymbol{\alpha})$) and spectral curvature  ($S_{\rm TV}(\boldsymbol{\alpha})$) in the objective function. In the left panel, we show the spectral index map at 230 GHz of the blurred simulation image.
In the second panel (``no TV''), we show the reconstruction with both spectral TV hyperparameters set to zero, $\lambda_{TV\alpha}=\lambda_{TV\beta}=0$.
In the third panel (``Fiducial TV''), we show the reconstruction with the spectral TV hyperparameters set to the fiducial values used in \autoref{sec:ngEHT86-345}, $\lambda_{TV\alpha}=20$, $\lambda_{TV\beta}=30$.
In the third panel (``High TV''), we show the reconstruction with large values of the the spectral TV hyperparameters,  $\lambda_{TV\alpha}=\lambda_{TV\beta}=100$. All other hyperparameters and settings in the image reconstruction were identical in each reconstruction.}
\label{fig:tvtest}
\end{figure*}

In \autoref{sec:ngEHT86-345}, we showed that while multi-frequency RML image reconstructions perform much better than single-frequency reconstructions in recovering accurate spectral index information from simulated ngEHT datasets of M87*. However, \autoref{fig:rjetspec} illustrates that multi-frequency RML image reconstructions may still suffer from artifacts in their spectral index maps. Specifically, the recovered spectral index map $\boldsymbol{\alpha}$ in \autoref{fig:rjetspec} features an inaccurate patch of high spectral index on the lower jet edge.

All image reconstructions from sparse VLBI datasets may contain errors or artificial features while still being good fits to the observations. In RML imaging, the presence of these features may be enhanced or mitigated by the choice of the hyperparameter values $\lambda_R$ in the  objective function (\autoref{eq::objfunc}). In \autoref{fig:tvtest}, we show three alternate reconstructions of the same data used in \autoref{fig:rjetspec} where the imaging procedure is identical except for the values of the hyperparameters $\lambda_{TV}$ weighting the weighting the total variation regularizer term for the spectral index ($S_{\rm TV}(\boldsymbol{\alpha})$) and spectral curvature ($S_{\rm TV}(\boldsymbol{\alpha})$). Using no total variation regularization results in many small-scale image artifacts in the spectral index map, while increasing the values of the hyperparameter $\lambda_{TV}$ suppresses small-scale structure and smooths out the spectral index map.

The fiducial hyperparameter values used in the text in \autoref{fig:rjetspec} remove most of the artificial features seen in the zero-regularizer case, except for the most prominent artifact at the jet edge. When we increase the value of the hyperparameters further in the high-TV case, this artifact is eliminated, but some real variations in the spectral index map across the extended jet are also suppressed.

Choosing hyperparameters in RML imaging is not trivial. In the main text, we selected fiducial hyperparameter weights $\lambda_R$ that worked well across a variety of synthetic data sets and were not fine-tuned to any particular image; we also chose values that were not too large relative to the data weights $\kappa_D$ in the objective function. In practice, it
is useful to survey over multiple combinations in the hyperparameter space before settling on final values and to apply hyperparameter values that work reasonably well on a large number of different synthetic data sets rather than fine-tuning the hyperparameters to perfectly reconstruct images from one example \citep{PaperIV}.
\clearpage
\bibliography{allrefs}{}
\bibliographystyle{aasjournal}

\end{document}